\documentclass[twocolumn,           % Format : preprint, twocolumn
               showpacs,            % Pacs : showpacs, noshowpacs
               nopreprintnumbers,     % Preprint: preprintnumbers,
               aps,                 % Society: ...
               prd,          	    % Journal Style : pra, prb, prc, prd, pre,
               letterpaper,             % Size : a4paper, ...
               groupeaddress,      % Affiliation (Title) : groupedaddress,
               nofootinbib,         % Footnote: footinbib, nofootinbib
               tightenlines,        % Remove additional spaces in a line
               floats,floatfix      % Floating pictures and tables
               ]{revtex4-1}

\usepackage{graphicx}% Include figure files
\usepackage{dcolumn}% Align table columns on decimal point
\usepackage{bm}% bold math
\usepackage{amsmath}
\usepackage{amsfonts,amssymb}
\usepackage{color}
\usepackage{enumerate}

\begin{document}

\title{Constraining brane tension using rotation curves of galaxies}

\author{Miguel A. Garc\'{\i}a-Aspeitia$^{1,2}$}
\email{aspeitia@fisica.uaz.edu.mx}

\author{Mario A. Rodr\'iguez-Meza$^{3}$}
\email{marioalberto.rodriguez@inin.gob.mx}

\affiliation{$^{1}$Consejo Nacional de Ciencia y Tecnolog\'ia, \\ Av. Insurgentes Sur 1582. Colonia Cr\'edito Constructor, Del. Benito Ju\'arez C.P. 03940, M\'exico D.F. M\'exico}
\affiliation{$^{2}$Unidad Acad\'emica de F\'isica, Universidad Aut\'onoma de Zacatecas, Calzada Solidaridad esquina con Paseo a la Bufa S/N C.P. 98060, Zacatecas, M\'exico}

\affiliation{$^{3}$Departamento de F\'\i sica, Instituto Nacional de Investigaciones Nucleares, Apdo. Postal 18-1027, M\'exico D.F. 11801, M\'exico}

\begin{abstract}
We present in this work a study of brane theory phenomenology focusing on the brane tension parameter, which is the main observable of the theory. 
We show the modifications steaming from the presence of branes in the rotation 
curves of spiral galaxies for
three well known dark matter density profiles:
 the Pseudo isothermal, Navarro-Frenk-White and Burkert dark matter density profiles. 
We estimate the brane tension parameter using a sample of high resolution observed rotation curves of low surface brightness spiral galaxies and a synthetic rotation curve for the three density profiles. 
Also, the fittings using the brane theory model of the rotation curves are compared with standard Newtonian models.
We found that Navarro-Frenk-White model prefers lower values of the brane tension parameter, 
on the average $\lambda \sim 0.73\times 10^{-3}$ eV$^4$, therefore showing clear brane effects. Burkert case does prefer higher values of the tension parameter, 
on the average $\lambda \sim 0.93$ eV$^4$ -- $46$ eV$^4$, i.e., negligible brane effects. Whereas pseudo isothermal is an intermediate case.
In this context, we found that our results show weaker bounds to the brane tension values in comparison with other bounds found previously.
\end{abstract}

\keywords{Brane theory, astrophysics, dark matter, rotation curves, spiral galaxies}
\draft
\pacs{04.50.-h,98.62.Dm}
\date{\today}
\maketitle

%%%%%%%%%%%%%%%%%%%%%%%%%%%%%%%%%%%%%%%%
\section{Introduction} \label{Int}
%%%%%%%%%%%%%%%%%%%%%%%%%%%%%%%%%%%%%%%%%

General Theory of Relativity (GR) is the cornerstone of
astrophysics and cosmology, giving predictions with unprecedented success. 
At astrophysical scales GR has been tested in, for example, the solar system, stellar dynamics, black hole formation and evolution, among others (see for instance\cite{FischbachTalmadge1999,Will1993,Kamionkowski:2007wv,*Matts,*Peebles:2013hla}). 
However, GR is being currently tested with various phenomena that can be 
significant challenges to the GR theory, generating important changes never seen before. Ones of the major challenges of modern cosmology are undoubtedly dark matter (DM) and dark energy. They comprise approximately $27\%$ for DM and $68\%$ for dark energy, of our 
universe\cite{PlanckCollaboration2013} allowing the formation of large scale structures\cite{FW,Diaferio:2008jy}.
Dark matter has been invoked as the mechanism to stabilize spiral galaxies and to provide with a matter distribution component to explain the observed rotation curves.
Nowadays the best model of the universe we have is the 
\emph{concordance} or $\Lambda$CDM model that has been successful in explaining the very large-scale structure formation, the statistics of the distribution of galaxy clusters, the temperature anisotropies of the cosmic microwave background radiation (CMB) and many other astronomical observations. 
In spite of all successes we have mentioned, 
this model has several problems, for example: predicts too much power on small scale\cite{Rodriguez-Meza:2012}, then it over predicts the number of observed satellite galaxies\cite{Klypin1999,Moore1999,Papastergis2011} and predicts halo profiles that are denser and 
cuspier than those inferred observationally\cite{Navarro1997,Subramanian2000,Salucci}, and also predicts a population of massive and concentrated subclass that are inconsistent 
with  observations of the kinematics of Milky Way satellites\cite{BoylanKolchin2012}.

One of the first astronomical observations that brought attention on DM was the observation of rotation curves of spiral galaxies by Rubin and coworkers\cite{Rubin2001}, 
these observations turned out to be the main tool to investigate the role of DM at galactic scales: its role in determining the structure, how the mass is distributed, and the dynamics, evolution, and formation of spiral galaxies.
Remarkably, the corresponding rotation velocities of galaxies, can be explained with the density profiles of different Newtonian DM models like Pseudo Isothermal profile (PISO)\cite{piso}, Navarro-Frenk-White profile (NFW)\cite{Navarro1997} 
or Burkert profile\cite{Burkert}, among 
others\cite{Einasto}; except by the fact that until now it is unsolved the problem of cusp and core in the densities profiles. In this way, none of them have the last word because the main questions, about the density distribution and of course the \emph{nature} of DM, has not been resolved.

Alternative theories of gravity have been used to model DM. For instance a scalar field has been proposed to model DM\cite{Dick1996,Cho/Keum:1998}, 
and has been used to study rotation curves of spiral galaxies\cite{Guzman/Matos:2000}. This scalar field is coupled minimally to the metric, however, scalar fields coupled non minimally to the metric have also been used to study DM\cite{Rodriguez-Meza:2012,RodriguezMeza/Others:2001,RodriguezMeza/CervantesCota:2004,Rodriguez-Meza:2012b}. Equivalently $F(R)$ models exists in the literature that analyzes rotation curves\cite{Martins/Salucci:2007}.

On the other hand, one of the best candidates to extend GR is the brane theory, whose main characteristic is to add another dimension having a five dimensional bulk where it is embedded a four dimensional manifold called the brane\cite{Randall-I,*Randall-II}. This model is characterized by the fact that the standard model of particles is confined in the brane and only the gravitational interaction can travel in the bulk\cite{Randall-I,*Randall-II}. The assumption that the five dimensional Einstein's equations are valid, generates corrections in the four dimensional Einstein's equations confined in the brane bringing information from the extra dimension\cite{sms}. 
These extra corrections in the Einstein's equations can help us to elucidate and solve the problems that afflicts the modern cosmology and astrophysics\cite{m2000,*yo2,*Casadio2012251,*jf1,*gm,*Garcia-Aspeitia:2013jea,*langlois2001large,*Garcia-Aspeitia:2014pna,*jf2,*PerezLorenzana:2005iv,*Ovalle:2014uwa,*Garcia-Aspeitia:2014jda,*Linares:2015fsa,*Casadio:2004nz}.

Before we start, let us mention here some experimental constraints on braneworld models, most of them about the so-called brane tension, $\lambda$, which appears explicitly as a free parameter in the corrections of the gravitational equations mentioned above. As a first example we have the measurements on the deviations from Newton's law of the gravitational interaction at small distances. It is reported that no deviation is observed for distances $l \gtrsim  0.1 \, {\rm mm}$, which then implies a lower limit on the brane tension in the Randall-Sundrum II model (RSII): $\lambda> 1 \, {\rm TeV}^{4}$\cite{Kapner:2006si,*Alexeyev:2015gja}; it is important to  mention that these limits do not apply to the two-branes case of the  Randall-Sundrum I model (RSI) (see \cite{mk} for details). 
Astrophysical studies, related to gravitational waves and stellar stability, constrain
the brane tension to be  $\lambda > 5\times10^{8} \, {\rm MeV}^{4}$\cite{gm,Sakstein:2014nfa}, whereas the existence of black hole X-ray binaries suggests that $l\lesssim 10^{-2} {\rm mm}$\cite{mk,Kudoh:2003xz,*Cavaglia:2002si}. Finally, from cosmological observations, the requirement of successful nucleosynthesis provides the lower limit $\lambda> 1\, {\rm MeV}^{4}$, which is a much weaker limit as compared to other experiments (another cosmological tests can be seen in: Ref. \cite{Holanda:2013doa,*Barrow:2001pi,*Brax:2003fv}).
  
In fact, this paper is devoted to study the main observable of brane theory which is the brane tension, whose existence delimits between the four dimensional GR and its high energy corrections. We are given to the task of perform a Newtonian approximation of the modified Tolman-Oppenheimer-Volkoff (TOV) equation, maintaining the effective terms provided by branes which cause subtle differences in the traditional dynamics. 
In this way we test the theory at galactic scale using high resolution measurements of rotation curves of a sample of low surface brightness (LSB) galaxies with no photometry\cite{deBlok/etal:2001}
and a synthetic rotation curve built from 40 rotation curves of spirals of magnitude around $M_I=-18.5$ where was found that the baryonic components has a very small contribution\cite{Salucci1},
assuming PISO, NFW and Burkert DM profiles respectively and with that, we constraint the preferred value of brane tension with observables. 
That the sample has no photometry means that the galaxies are DM dominated and then we have only two parameters related to the distribution of DM, a density scale and a length scale, adding the brane tension we have three parameters in total to fit. 
The brane tension fitted values are compared among the traditional DM density profile models of spiral galaxies 
(PISO, NFW and Burkert) and against the same models without the presence of branes and confronted with other values of the tension parameter coming from cosmological and astrophysical observational data.

This paper is organized as follows: In Sec.\ \ref{EM} we show the equations of motion (modified TOV equations) for a spherical symmetry and the appropriate initial conditions. In Sec.\ \ref{TOV MOD} we explore the Newtonian limit and we show the mathematical expression of rotation velocity with brane modifications; particularly we show the modifications to velocity rotation expressions of PISO, NFW and Burkert DM profiles and they are compared with models without branes. 
In Sec.\ \ref{Results} we test the DM models plus brane with observations: we use a sample of high resolution measurements of rotation curves of LSB galaxies and a synthetic rotation curve representative of  40 rotation curves of spirals where the baryonic component has a very small contribution.
Finally in Sec.\ \ref{Disc}, we discuss the results obtained in the paper and make some conclusions.

In what follows, we work in units in which $c=\hbar=1$, unless explicitly written.

%%%%%%%%%%%%%%%%%%%%%%%%%%%%%%%%%%%%%%%%%
\section{Review of equations of motion for branes} \label{EM}
%%%%%%%%%%%%%%%%%%%%%%%%%%%%%%%%%%%%%%%%%

Let us start by writing the equations of motion for galactic stability in a brane embedded in a five-dimensional bulk according to the RSII model\cite{Randall-II}. Following an appropriate computation (for details see\cite{mk,sms}), it is possible to demonstrate that the modified four-dimensional Einstein's equations can be written as 
\begin{equation}
  G_{\mu\nu} + \xi_{\mu\nu} + \Lambda_{(4)}g_{\mu\nu} = \kappa^{2}_{(4)} T_{\mu\nu} + \kappa^{4}_{(5)} \Pi_{\mu\nu} +
  \kappa^{2}_{(5)} F_{\mu\nu} , \label{Eins}
\end{equation}
where $\kappa_{(4)}$ and $\kappa_{(5)}$ are respectively the four and five- dimensional coupling constants, which are related in the form: $\kappa^{2}_{(4)}=8\pi G_{N}=\kappa^{4}_{(5)} \lambda/6$, where $\lambda$ is defined as the brane tension, and $G_{N}$ is the Newton constant. For purposes of simplicity, we will not consider bulk matter, which translates into $F_{\mu\nu}=0$, and discard the presence of the four-dimensional cosmological constant, $\Lambda_{(4)}=0$, 
as we do not expect it to have any important effect at galactic scales (for a recent discussion about it see\cite{Pavlidou:2013zha}). Additionally, we will neglect any nonlocal energy flux, which is allowed by the static spherically symmetric solutions we will study below\cite{gm}.

The energy-momentum tensor, the quadratic energy-momentum tensor, and the Weyl (traceless) contribution, have the explicit forms
\begin{subequations}
\label{eq:4}
\begin{eqnarray}
\label{Tmunu}
T_{\mu\nu} &=& \rho u_{\mu}u_{\nu} + p h_{\mu\nu} \, , \\
\label{Pimunu}
\Pi_{\mu\nu} &=& \frac{1}{12} \rho \left[ \rho u_{\mu}u_{\nu} + (\rho+2p) h_{\mu\nu} \right] \, , \\
\label{ximunu}
\xi_{\mu\nu} &=& - \frac{\kappa^4_{(5)}}{\kappa^4_{(4)}} \left[ \mathcal{U} u_{\mu}u_{\nu} + \mathcal{P}r_{\mu}r_{\nu}+ \frac{ h_{\mu\nu} }{3} (\mathcal{U}-\mathcal{P} ) \right] \, .
\end{eqnarray}
\end{subequations}
Here, $p$ and $\rho$ are, respectively, the pressure and energy density of the stellar matter of interest, $\mathcal{U}$ is the nonlocal energy density, and $\mathcal{P}$ is the nonlocal anisotropic stress. Also, $u_{\alpha}$ is the four-velocity (that also satisfies the condition $g_{\mu\nu}u^{\mu}u^{\nu}=-1$), $r_{\mu}$ is a unit radial vector, and $h_{\mu\nu} = g_{\mu\nu} + u_{\mu} u_{\nu}$ is the projection operator orthogonal to $u_{\mu}$.

Spherical symmetry indicates that the metric can be written as:
\begin{equation}
{ds}^{2}= - B(r){dt}^{2} + A(r){dr}^{2} + r^{2} (d\theta^{2} + \sin^{2} \theta d\varphi^{2}) \, .\label{metric}
\end{equation}
If we define the reduced Weyl functions $\mathcal{V} = 6 \mathcal{U}/\kappa^4_{(4)}$, and $\mathcal{N} = 4 \mathcal{P}/\kappa^4_{(4)}$. First, we define the effective mass as:
\begin{equation}
\mathcal{M}^\prime_{eff} = 4\pi{r}^{2}\rho_{eff}. \label{eq:7a}
\end{equation}
Then, from Eqs. \eqref{Eins} and \eqref{eq:4} and after straightforward calculations we have the following equations of motion:
\begin{subequations}
  \label{eq:7}
\begin{eqnarray}
  p^\prime &=& -\frac{G_N}{r^{2}} \frac{4 \pi \, p_{eff} \, r^3 + \mathcal{M}_{eff}}{1 - 2G_N \mathcal{M}_{eff}/r} ( p + \rho ) \, ,   \label{eq:7b} \\
  \mathcal{V}^{\prime} + 3 \mathcal{N}^{\prime} &=& - \frac{2G_N}{r^{2}} \frac{4 \pi \, p_{eff} \, r^3 + \mathcal{M}_{eff}}{1 - 2G_N \mathcal{M}_{eff}/r} \left( 2 \mathcal{V} + 3 \mathcal{N} \right)\nonumber\\ 
  && - \frac{9}{r} \mathcal{N} - 3 (\rho+p) \rho^{\prime} \, ,  \label{eq:7c}
\end{eqnarray}
\end{subequations}
where a prime indicates derivative with respect to $r$, $A(r) = [1 - 2G_N \mathcal{M}(r)_{eff}/r]^{-1}$, and the effective energy density and pressure, respectively, are given as:
\begin{subequations}
\label{eq:3}
\begin{eqnarray}
\rho_{eff}  &=& \rho \left( 1 + \frac{\rho}{2\lambda} \right) + \frac{\mathcal{V}}{\lambda}  \, , \label{eq:3a} \\
p_{eff} &=& p \left(1 + \frac{\rho}{\lambda} \right) + \frac{\rho^{2}}{2\lambda} + \frac{\mathcal{V}}{3\lambda} + \frac{\mathcal{N}}{\lambda} \, . \label{eq:3b}
\end{eqnarray}
\end{subequations}
Even though we will not consider exterior galaxy solutions, we must anyway take into account the information provided by the Israel-Darmois (ID) matching condition, which for the case under study can be written as\cite{gm}:
\begin{equation}
  \label{eq:28}
    (3/2) \rho^2(R) + \mathcal{V}^-(R) + 3 \mathcal{N}^-(R) = 0 \, .
\end{equation}
In this case, the superscript ($-$) indicates the interior value of the quantity at the halo surface\footnote{We denote the surface of the galaxy as a region where does not exist DM or baryons, \emph{i.e.}, the intergalactic space.} of the galaxy, assuming that $\rho(r>R)=0$ where $R$ denotes the maximum size of the galaxy. Also, the previous equation takes in consideration the fact that the exterior must be Schwarzschild which in general the following condition must be fulfilled $\mathcal{V}(r \geq R) = 0 =\mathcal{N}(r\geq R)$ (see\cite{Garcia-Aspeitia:2014pna} for details).

For completeness, we just note that the exterior solutions of the metric functions are given by the well known expressions $B(r) = A^{-1}(r) = 1 - 2G_N M_{eff}/r$.

Finally, we impose $\mathcal{N}=0$ (see\cite{Garcia-Aspeitia:2014pna}). Implying that Eq. \eqref{eq:28} is restricted as:
\begin{equation}
  \label{eq:29}
    -(3/2) \rho^2(R) = \mathcal{V}^-(R) \, ,
\end{equation}
with the aim of maintain a galaxy Schwarzschild exterior.

%%%%%%%%%%%%%%%%%%%%%%%%%%%%%%%%%%%%%%%%
\section{Low energy limit and rotation curves} \label{TOV MOD}
%%%%%%%%%%%%%%%%%%%%%%%%%%%%%%%%%%%%%%%%

To begin with, we observe, from Eq.\ \eqref{eq:7b} in the low energy (Newtonian) limit, that we have: $r^{2}p^{\prime}=-G_{N}\mathcal{M}_{eff}\rho$. Differentiating we found
\begin{equation}
\frac{d}{dr}\left(\frac{r^{2}}{\rho}\frac{dp}{dr}\right)=-4\pi r^{2}G_{N}\rho_{\rm eff}. \label{eqdiff9}
\end{equation}
From here it is possible to note that $d\Phi/dr=-\rho^{-1}(dp/dr)$ resulting in
\begin{equation}
\nabla^{2}\Phi_{\rm eff}=\frac{1}{r^{2}}\frac{d}{dr}\left(r^{2}\frac{d\Phi_{\rm eff}}{dr}\right)=4\pi G_{N}\rho_{\rm eff}, \label{Poisson}
\end{equation}
being necessary to define the energy density of DM together with the nonlocal energy density. Notice that the nonlocal energy density can be obtained easily from Eq.\ \eqref{eq:7c} in the galaxy interior and also the fluid behaves like dust, implying the condition $p=0$, always fulfilling the low energy condition $4\pi r^3p_{eff}\ll\mathcal{M}_{eff}$ and $2G_{N}\mathcal{M}_{eff}/r\ll1$, between effective quantities and in consequence $4G_{N}\mathcal{M}_{eff}\mathcal{V}/r^2\sim0$, is negligible.

In addition, the rotation curve is obtained from the contribution of the effective potential, this expression can be written as:
\begin{eqnarray}
V^2(r) &=& r\left\vert\frac{d\Phi_{\rm eff}}{dr}\right\vert=\frac{G_N \mathcal{M}_{eff}(r)}{r} 
\nonumber \\
&=& 
\frac{G_N }{r} 
\left[
\mathcal{M}_{DM}(r) + \mathcal{M}_{Brane}(r)
\right]
, \label{rotvel}
\end{eqnarray}
where $\mathcal{M}_{DM}(r)$ is the contribution to the mass from DM, $\mathcal{M}_{Brane}(r)$ gives the modification to the DM mass that comes from the brane; and $\mathcal{M}_{eff}(r)$ must be greater than zero. From here, it is possible to study the rotation velocities of the DM, assuming a variety of density profiles.

Before we start let us define the following dimensionless variables: $\bar{r}\equiv r/r_{\rm s}$, $v_{0}^{2}\equiv4\pi G_{N}r_{\rm s}^{2}\rho_{\rm s}$ and $\bar{\rho}\equiv\rho_{\rm s}/2\lambda$ where $\rho_{\rm s}$, is the central density of the halo and $r_{s}$ is associated with the central radius of the halo. 

%%%%%%%%%%%%%%%%%%%%%%%%%%%%%%%
\subsection{Pseudo isothermal profile for dark matter}
%%%%%%%%%%%%%%%%%%%%%%%%%%%%%%%

Here we consider that DM density profile is given by PISO\cite{piso} written as:
\begin{equation}
\rho_{\rm PISO}(r)=\frac{\rho_{\rm s}}{1+\bar{r}^{2}}. \label{PIP}
\end{equation}
From Eq. \eqref{rotvel}, together with Eq. \eqref{PIP}, it is possible to obtain:
\begin{eqnarray}
V_{\rm PISO}^{2}(\bar{r}) &=& v_{0}^{2}
\left\lbrace
\left(
1-\frac{1}{\bar{r}}\arctan\bar{r}
\right) 
\right.
\nonumber \\
&& 
\left. + \bar{\rho}
\left(
\frac{1}{1+\bar{r}^2}- \frac{1}{\bar{r}}\arctan\bar{r}
\right)
\right\rbrace.
\label{RCPISO}
\end{eqnarray}
In the limit $\bar{\rho}\to0$, we recover the classical rotation velocity associated with PISO for DM.
The effective density must be positive defined, then $\lambda > \rho_s$ must be fulfilled. The first right-hand term in parenthesis in Eq.\ \eqref{RCPISO} is PISO dark matter contribution and the second 
is brane's contribution.

%%%%%%%%%%%%%%%%%%%%%%%%%%%%%%%
\subsection{Navarro-Frenk-White profile for dark matter}
%%%%%%%%%%%%%%%%%%%%%%%%%%%%%%%
Another interesting case (motivated by cosmological $N$-body simulations) is the NFW density profile, which is given by\cite{NFW}:
\begin{equation}
\rho_{\rm NFW}(r)=\frac{\rho_{\rm s}}{\bar{r}(1+\bar{r})^{2}}. \label{NFW}
\end{equation}
This is a density profile that diverges as $r \rightarrow 0$ 
and 
%not necessarily 
it is not possible to say
that $\rho_s$ is related with the central density of the DM distribution.
Also density goes as $1/\bar{r}^3$ when $\bar{r} \gg 1$.
However, in this particular case, we will still be calling them the \emph{central} density and radius of the NFW matter distribution.
From Eq.\ \eqref{rotvel}, together with Eq.\ \eqref{NFW} we obtain the following rotation curve:
\begin{eqnarray}
V_{\rm NFW}^{2}(\bar{r}) &=& v_{0}^{2}\left\lbrace\left(\frac{(1+\bar{r})\ln(1+\bar{r})-\bar{r}}{\bar{r}(1+\bar{r})}\right)\right.\nonumber\\&&
\left.+\frac{2\bar{\rho}}{3\bar{r}}\left(\frac{1}{(1+\bar{r})^{3}}-1\right) \right\rbrace.
\label{RCNFW}
\end{eqnarray}
The first right-hand term in parenthesis in Eq.\ \eqref{RCNFW} is NFW dark matter contribution and the second one is the brane's contribution. Notice that we recover also the classical limit when $\bar{\rho}\to0$.

In addition, it is important to remark that the effective density must be positive defined, then $\lambda > \rho_s r_s /r$. Also, if  $\mathcal{M}(r)$ must be greater than zero, then $r > r_{min}$ where $r_{min}$ is given by solving the following equation:
\begin{equation}
\frac{2}{3}\bar{\rho}=\frac{(\alpha+1)^2[(\alpha+1)\ln(\alpha+1)-\alpha]}{(\alpha+1)^3-1},\label{comp}
\end{equation}
where we define $\alpha\equiv r_{min}/r_s$ as a dimensionless quantity.

%%%%%%%%%%%%%%%%%%%%%%%%%%%%%%%
\subsection{Burkert density profile for dark matter}
%%%%%%%%%%%%%%%%%%%%%%%%%%%%%%%

Another density profile was proposed by Burkert\cite{Burkert}, which it has the form:
\begin{equation}
\rho_{\rm Burk}=\frac{\rho_{\rm s}}{(1+\bar{r})(1+\bar{r}^{2})}. \label{Burk}
\end{equation}
Again,  from Eq.\ \eqref{rotvel}, together with Eq.\ \eqref{Burk} we obtain the following rotation curve:
\begin{eqnarray}
V_{\rm Burk}^{2}(\bar{r}) &=&\frac{v_{0}^{2}}{4\bar{r}} \left\lbrace \left( \ln[(1+\bar{r}^{2})(1+\bar{r})^{2}]-2\arctan(\bar{r}) \right)\right.
\label{RCBurkert}
\\&&
\left.+ \frac{1}{2}\bar{\rho}\left( \frac{1}{1+\bar{r}}+\frac{1}{1+\bar{r}^{2}}+\arctan(\bar{r})-2 \right)\right\rbrace.
\nonumber
\end{eqnarray}
In the limit $\bar{\rho}\to0$, we recover the classical rotation velocity associated with Burkert density profile\cite{Burkert}.
The effective density must be positive defined, then $\lambda > \rho_s$. 
Again the first right-hand term in parenthesis in Eq.\ \eqref{RCBurkert} is
Burkert DM contribution and the second one comes from the
 brane's contribution.

%%%%%%%%%%%%%%%%%%%%%%%%%%%%%%%%%%%%%%%%%%%%%%%
\section{Constrictions with galaxies without photometry} \label{Results}
%%%%%%%%%%%%%%%%%%%%%%%%%%%%%%%%%%%%%%%%%%%%%%%

To start with the analysis, we $\chi^{2}$ best fit the observational rotation curves of the sample with:
\begin{equation}
\chi^{2}=\sum_{i=1}^{N}\left(\frac{V_{theo}-V_{exp \; i}}{\delta V_{exp\; i}}\right)^{2},
\label{chi2Eq}
\end{equation}
where $i$ runs from one up to the number of points in the data, $N$; $V_{theo}$, is computed according to the velocity profile under consideration 
and $\delta V_{exp\; i}$, is the error in the measurement of the rotational velocity. Notice that 
the free parameters are only for DM-Branes: $r_{s}$, $\rho_{s}$ and $\lambda$. 
In the tables below we show $\chi_{red}^{2} \equiv \chi^{2}/(N - n_p -1)$ where $n_p$ is the number of parameters to fit, being in our case, $n_p=3$.

The analyzed sample of galaxies are twelve high resolution rotation curves of LSB galaxies with no photometry (visible components, such as gas and stars, are negligible) as given in Ref.\cite{deBlok/etal:2001}. This sample was used to study DM equation of state (EoS) in Ref.\cite{Barranco/etal:2015}. We remark that in this part we use units such that $4 \pi G_{N}=1$, velocities are in km/s, and distances are given in kpc.

%%%%%%%%%%%%%%%%%%%%%%%%%%%%%%%%%%%%%%%%%%%%%%%
\subsection{Results: PISO profile + Branes}
%%%%%%%%%%%%%%%%%%%%%%%%%%%%%%%%%%%%%%%%%%%%%%%

We have estimated the parameters of the PISO+branes model 
and were compared with PISO model without brane contribution, minimizing the appropriate $\chi^2$ for the sample of observed rotational curves, using Eq.\ (\ref{chi2Eq}) with Eq.\ (\ref{RCPISO}) and taking into account that $\lambda > \rho_s$ must be fulfilled. 

\begin{figure}
\includegraphics[scale=0.33]{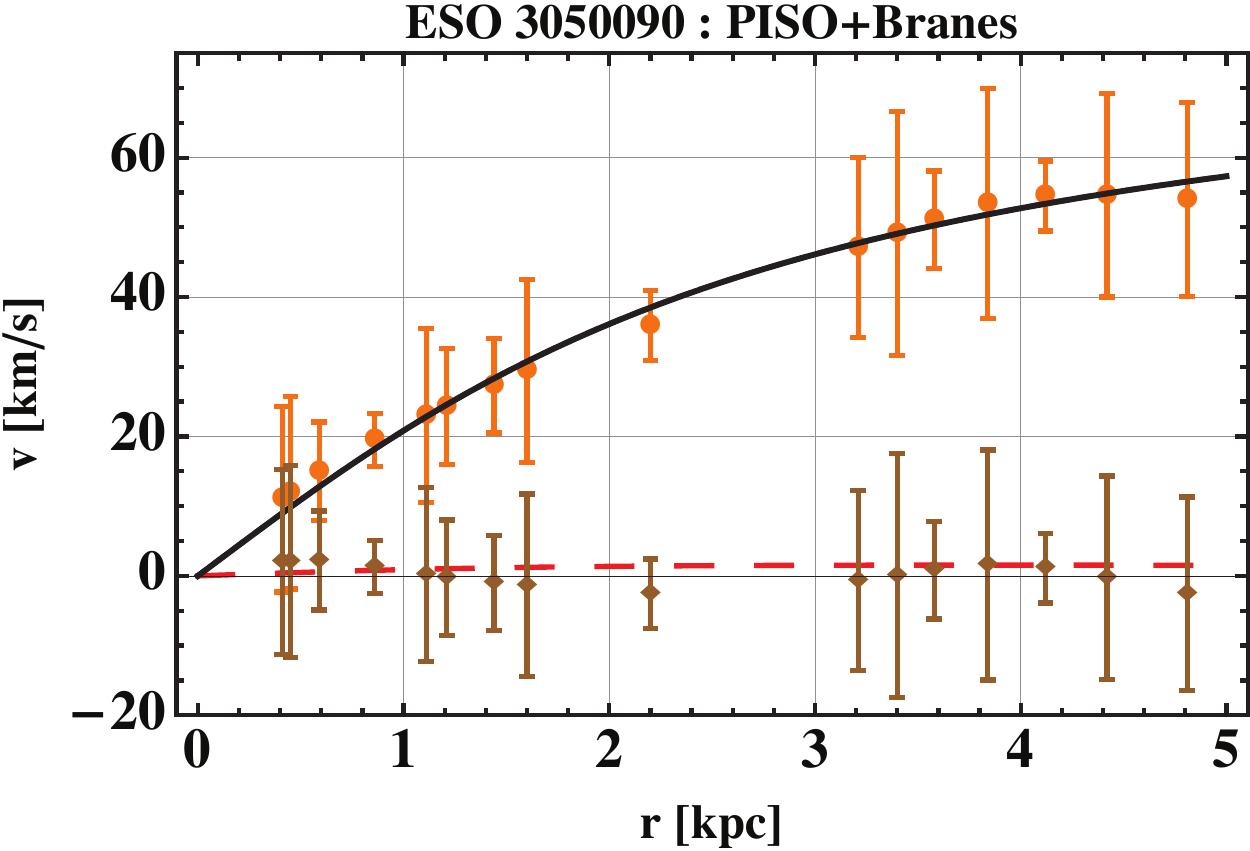}
\includegraphics[scale=0.33]{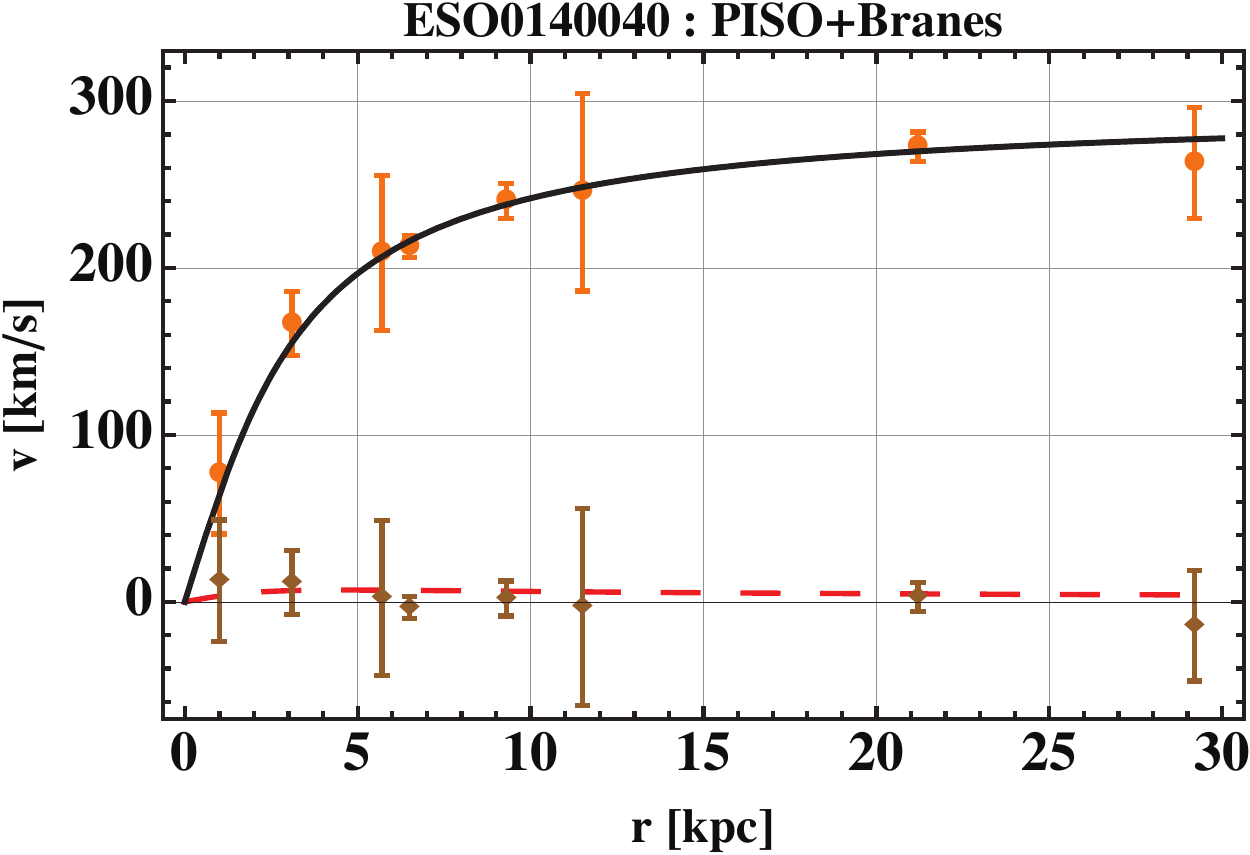} \\
\includegraphics[scale=0.33]{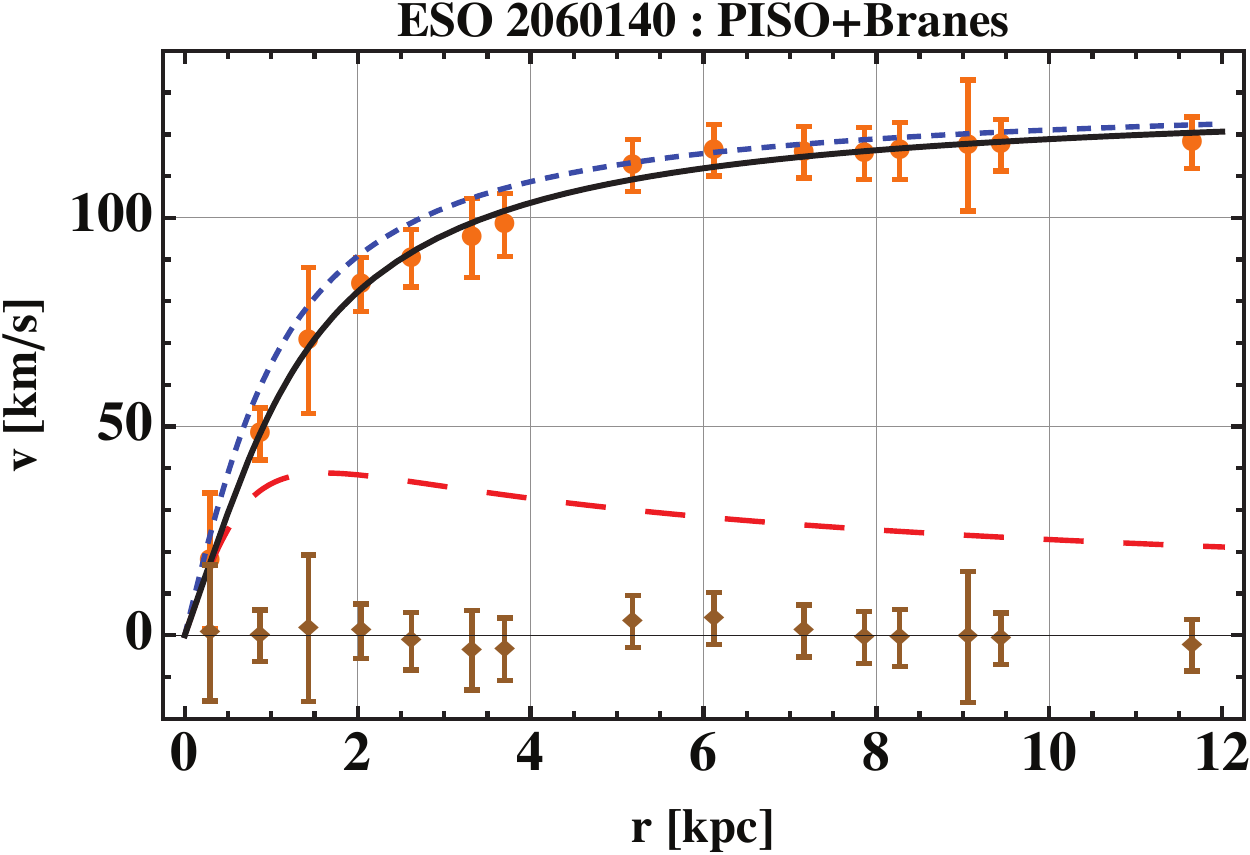}  
\includegraphics[scale=0.33]{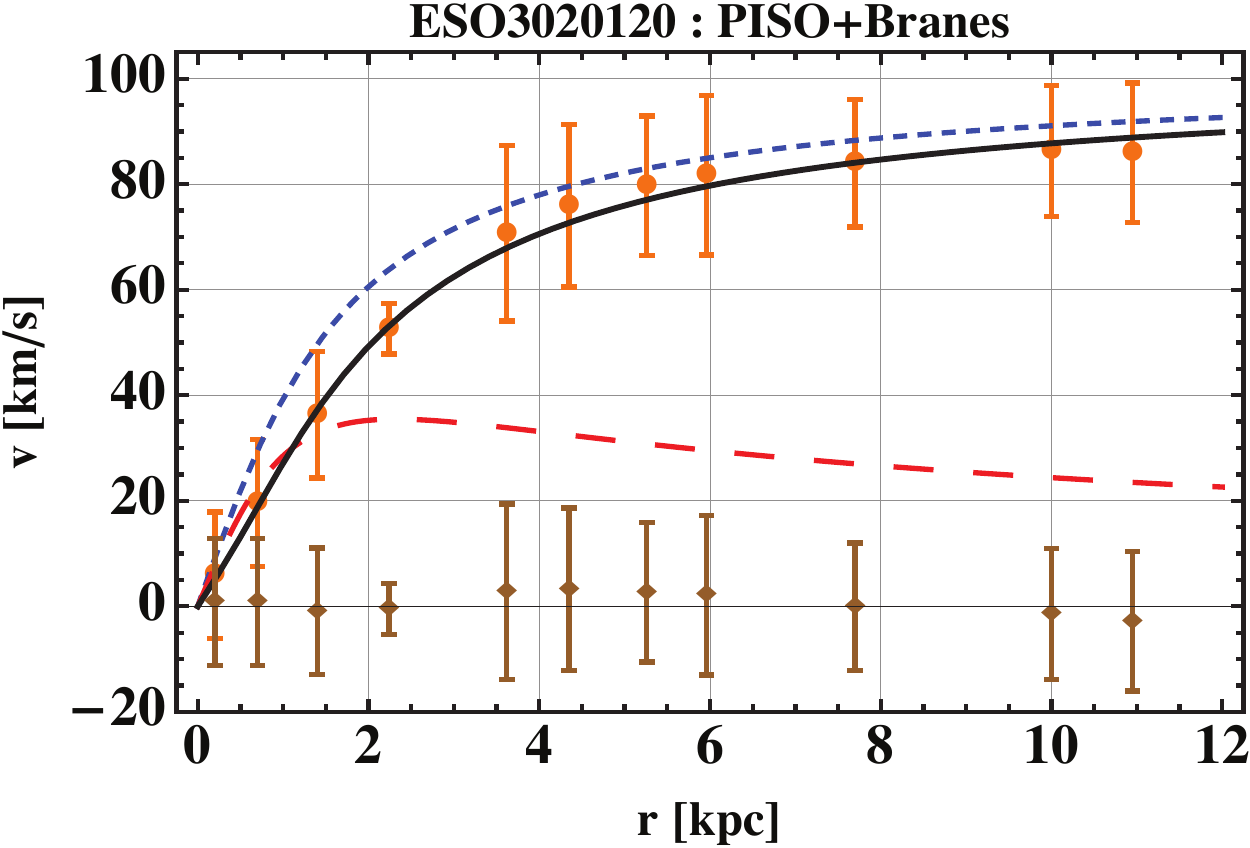} \\  
\includegraphics[scale=0.33]{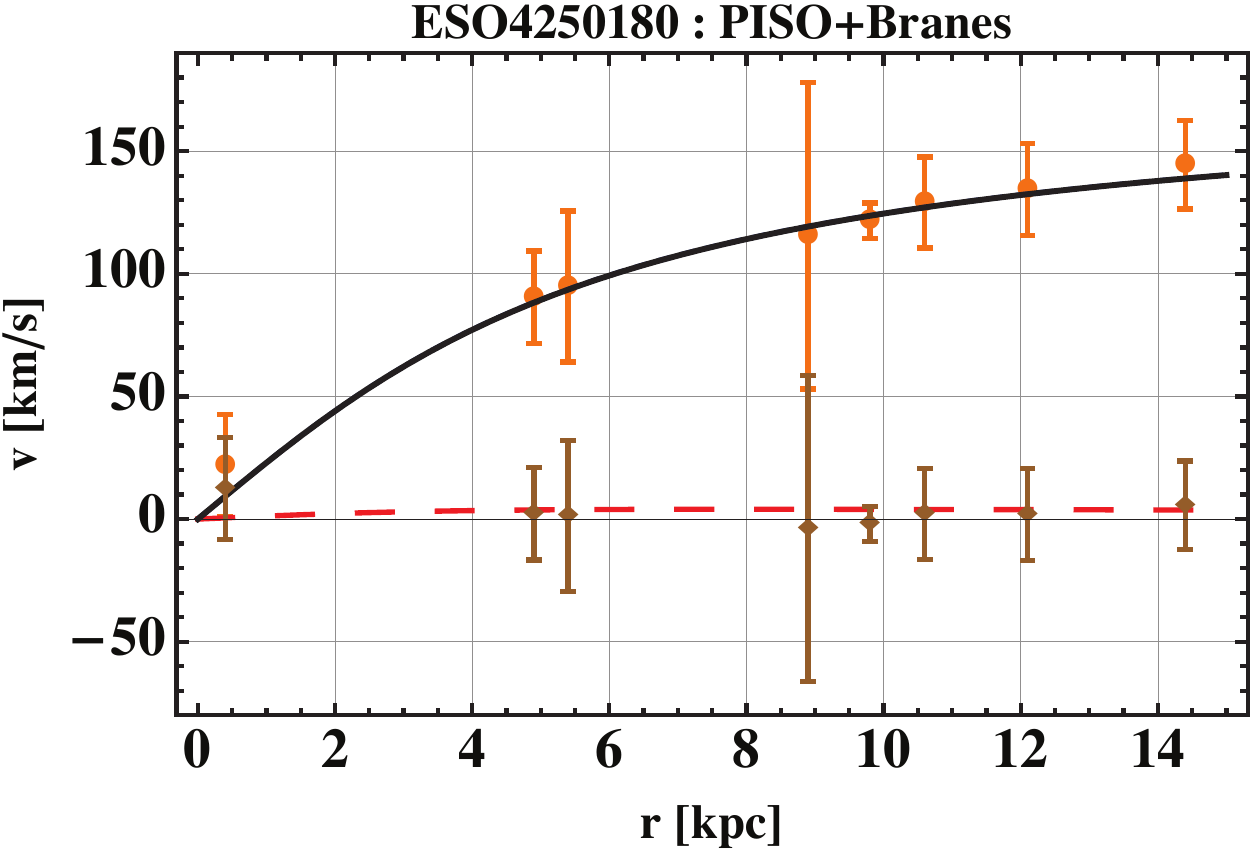} 
\includegraphics[scale=0.33]{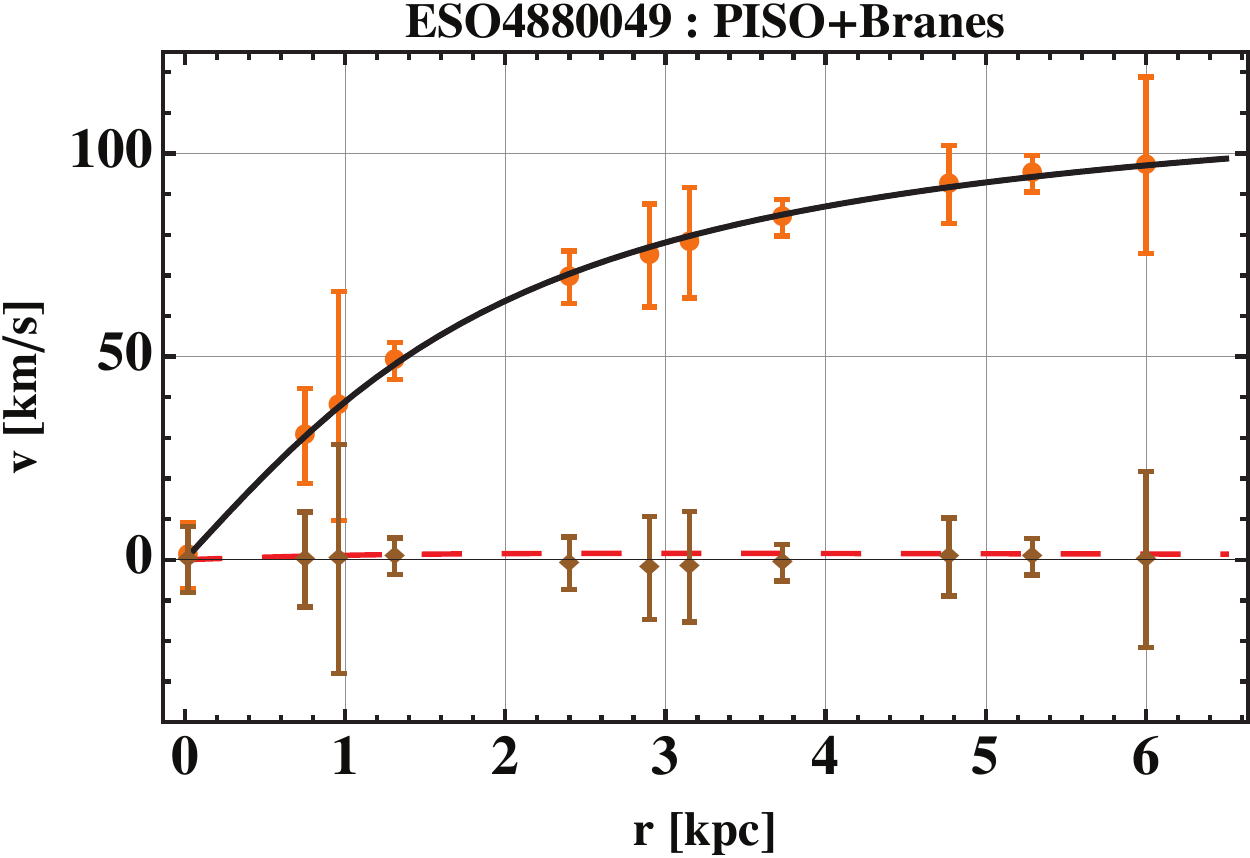} \\
\includegraphics[scale=0.33]{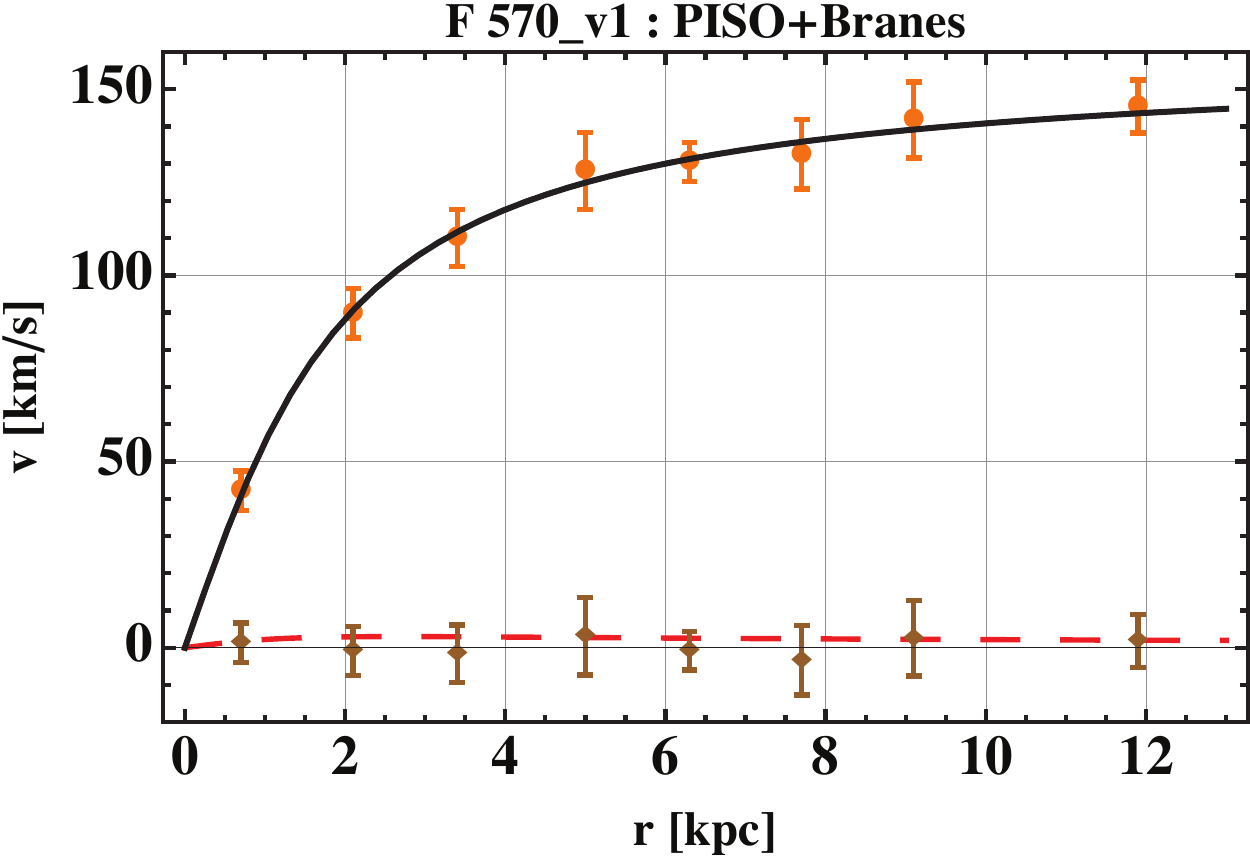} 
\includegraphics[scale=0.33]{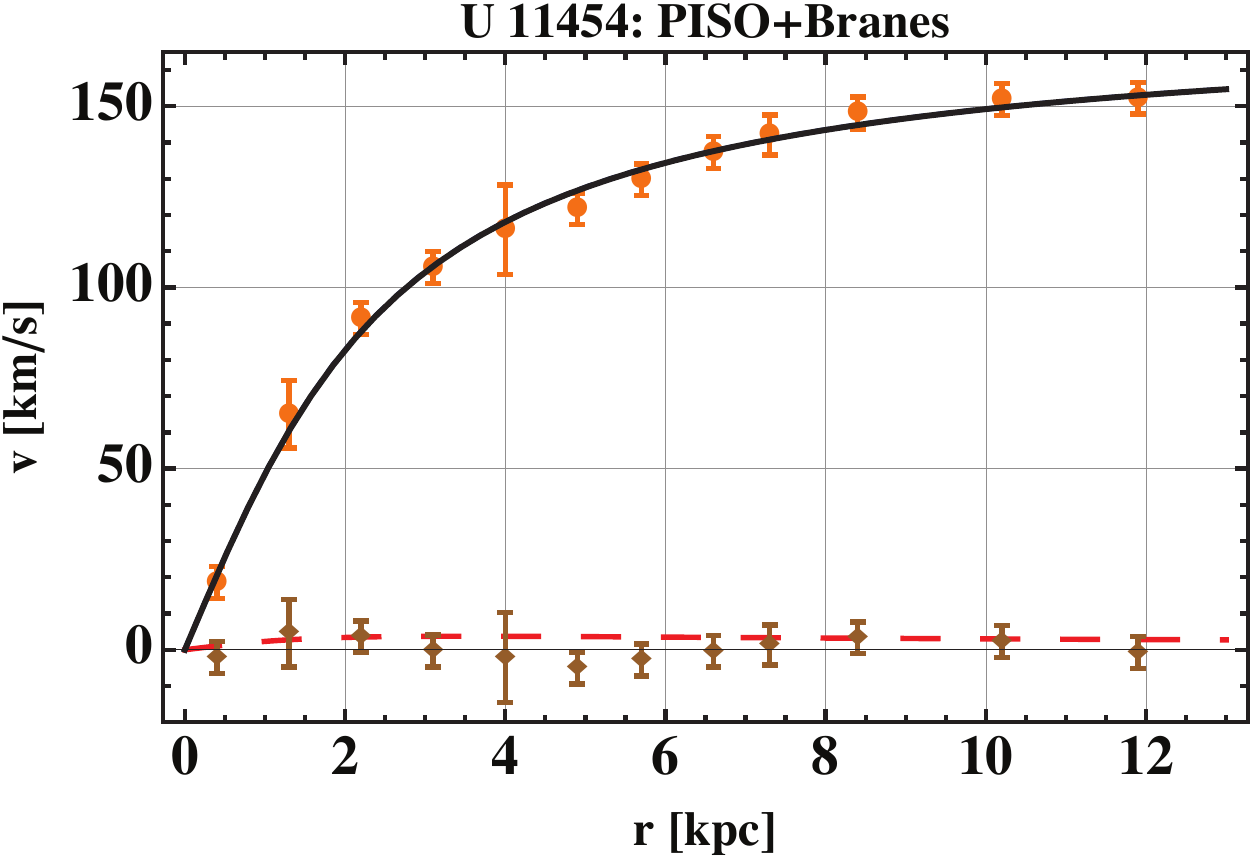} \\
\includegraphics[scale=0.33]{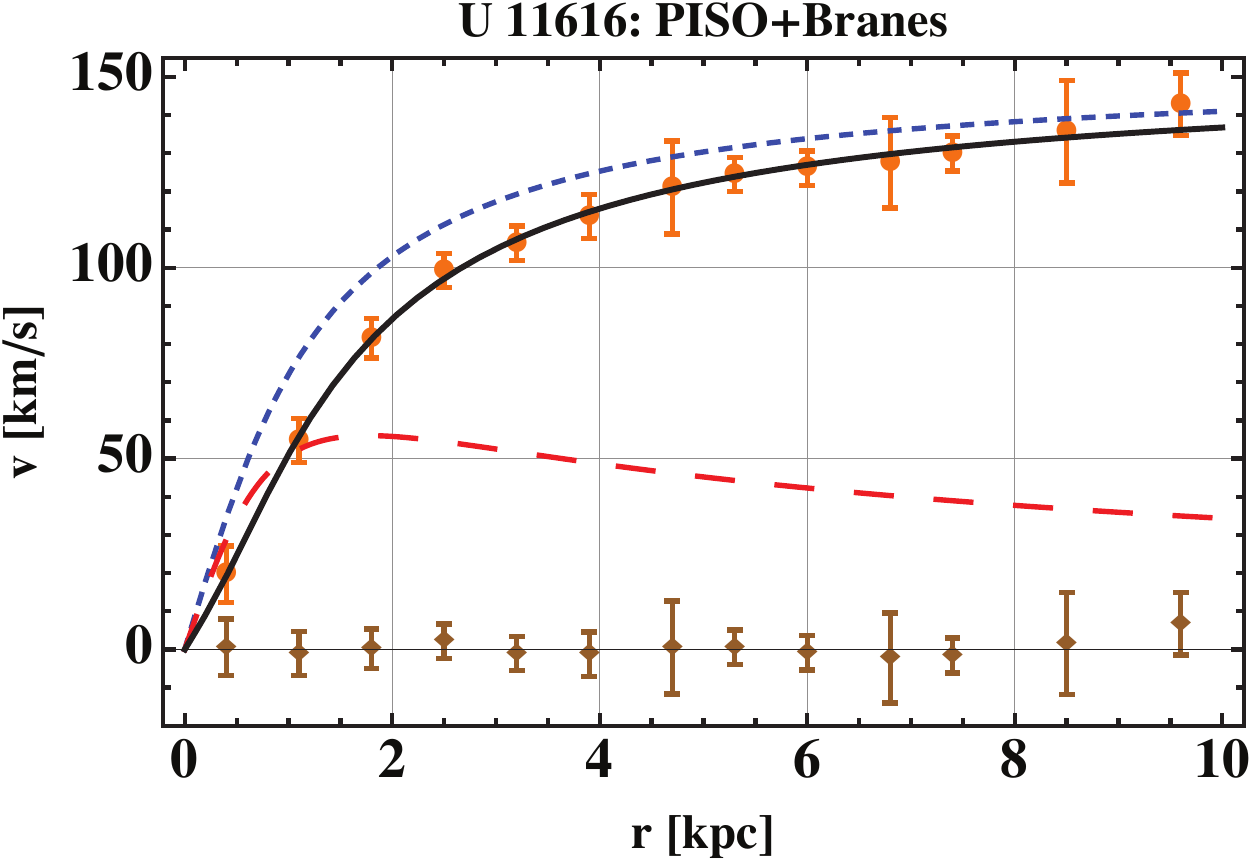} 
\includegraphics[scale=0.33]{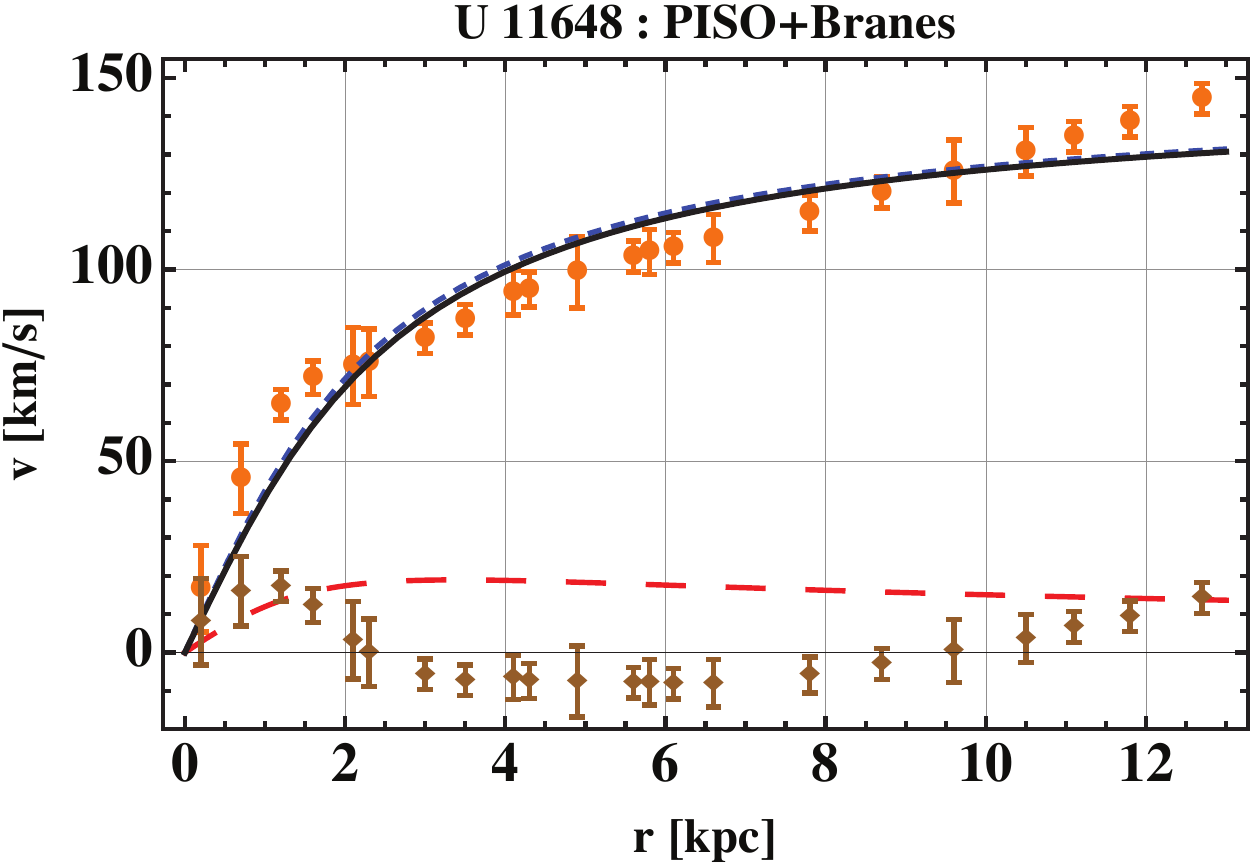} \\
\includegraphics[scale=0.33]{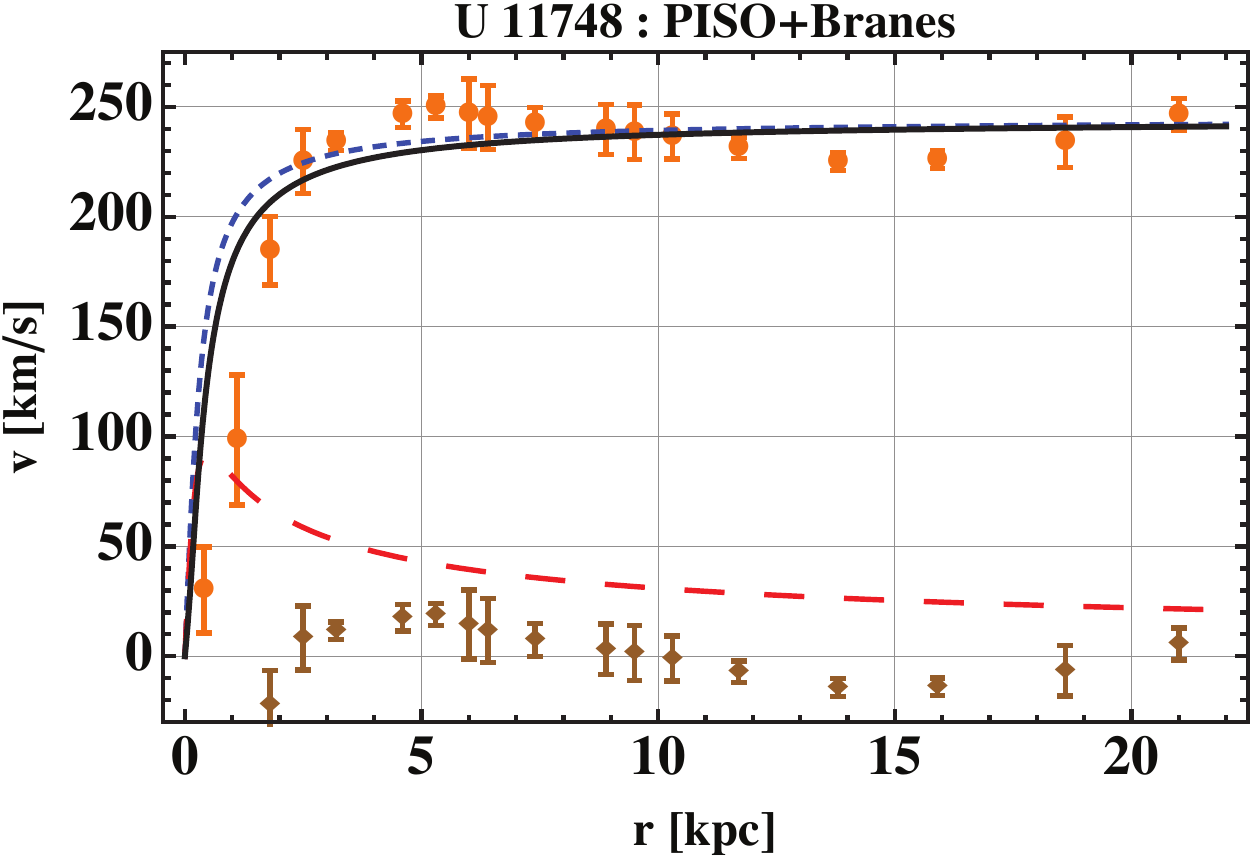} 
\includegraphics[scale=0.33]{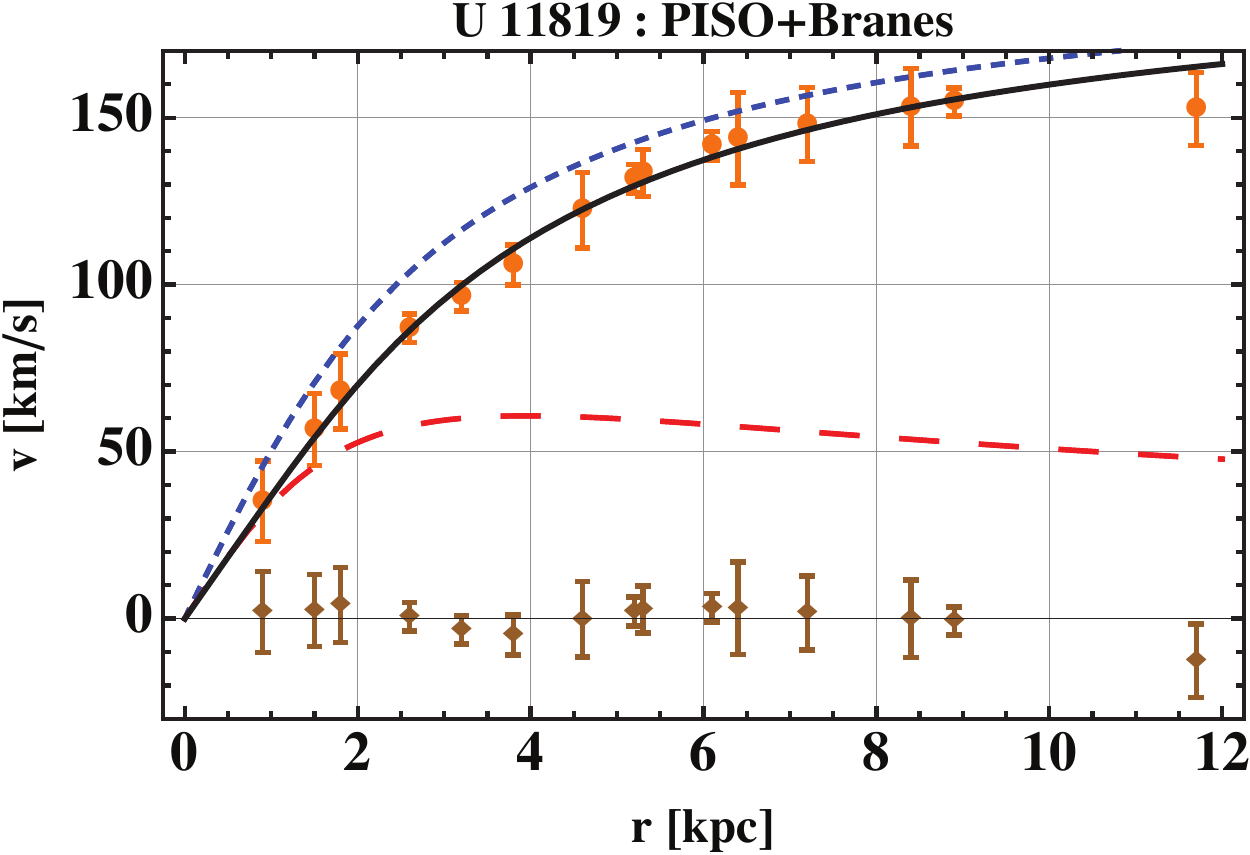}
\caption{Group of analyzed galaxies using modified rotation velocity for PISO profile: ESO 3050090,
ESO 0140040, ESO 2060140, ESO 3020120, ESO 4250180, ESO 4880049, 570\_V1, U11454, U11616, U11648, U11748, U11819. We show in the plots: Total rotation curve (solid black line), only PISO curve (short dashed blue curve) and the rotation curve associated with the mass lost by the effect of the brane (red dashed curve).} 
\label{PISO1}
\end{figure}

In Fig.\ \ref{PISO1} we show, for each one of the galaxies in the sample,
the plots of the PISO theoretical rotation curve (solid line), that best fit of the corresponding observational data (orange symbols); also shown are the errors of the estimation (brown symbols). 
For each galaxy we have plotted the contribution to the rotation velocity due only to the brane (red long-dashed curve) and only to the dark matter PISO density profile (blue short-dashed curve), see Eq.\ (\ref{RCPISO}).
Brane effects are very clear in galaxies: 
ESO 2060140,
ESO 3020120,
U 11616,
U 11648,
U 11748,
U 11819.
In Table \ref{TablePiso} it is shown the central density, central radius and the brane tension which is the free parameter of the brane theory (only in PISO+branes). As a comparison, it is also shown the central density and radius without brane contribution.
The worst fitted galaxies were (high $\chi_{red}^2$ value): 
U 11648,
U 11748.
The fitted brane tension values presents great dispersion, from the lower value: 
$0.167\; M_{\odot}/\rm pc^3$, ESO 3020120 to the higher value:
$108.096\; M_{\odot}/\rm pc^3$, ESO 4880049.
It is useful to have $\lambda$ in eV, where the conversion from solar masses to eV is: $1 M_{\odot}/\rm pc^3 \sim 2.915\times10^{-4}eV^4$. 
The brane tension parameter has an average value of $\langle\lambda\rangle_{\rm PISO} = 33.178 \; M_{\odot}/\rm pc^3$ with a standard deviation $\sigma_{\rm PISO} = 40.935 \; M_{\odot}/\rm pc^3$. Notice that we can't see a clear tendency to a $\lambda$ value or range of values.

%%%%%%%%%%%%%%%%%%%%%%%%%%%
% ORIGINAL POSITION OF PISO TABLE
%%%%%%%%%%%%%%%%%%%%%%%%%%%

%%%%%%%%%%%%%%%%%%%%%%%%%%%%%%%%%%%%%%%%%%%%%%%
\subsection{Results: NFW profile + Branes}
%%%%%%%%%%%%%%%%%%%%%%%%%%%%%%%%%%%%%%%%%%%%%%%
For the NFW density profile case we have the following results:
We have estimated  parameters with and without brane contribution by
minimizing the corresponding $\chi^2$, Eq.\ (\ref{chi2Eq}) with Eq.\ (\ref{RCNFW}), for the sample of observed rotation curves 
and taking into account that
$\lambda > \rho_s r_s /r$ in order to have an effective density positive defined, always fulfilling Eq.\ (\ref{comp}).

\begin{figure}
\includegraphics[scale=0.33]{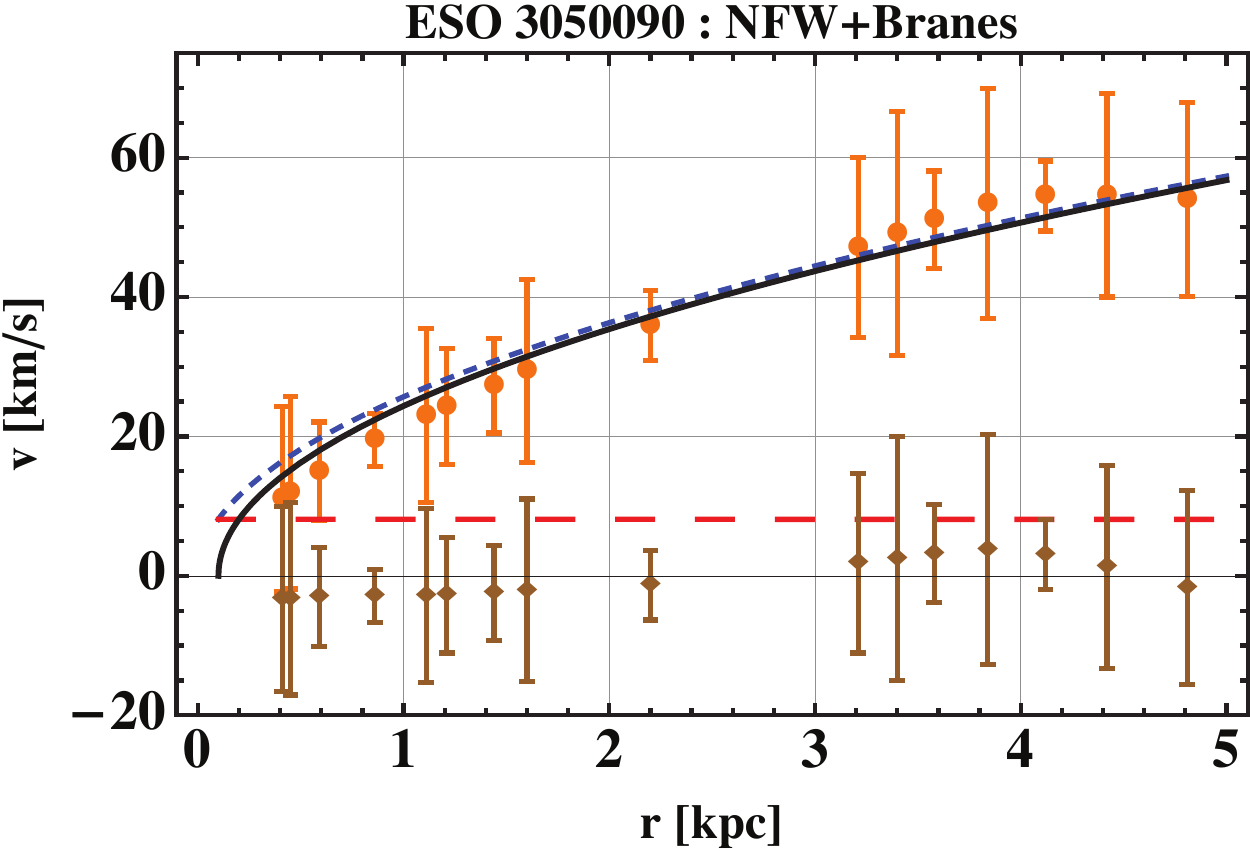} 
\includegraphics[scale=0.33]{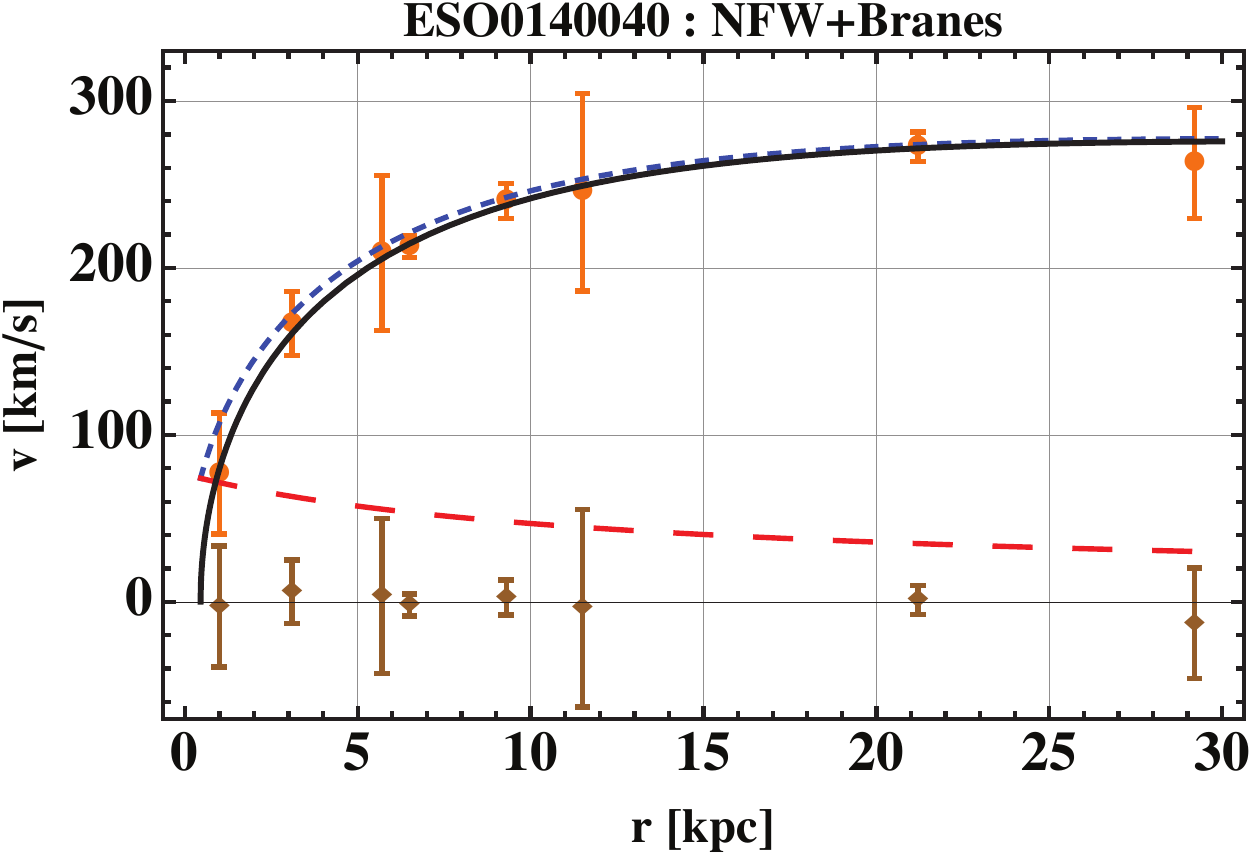} \\
\includegraphics[scale=0.33]{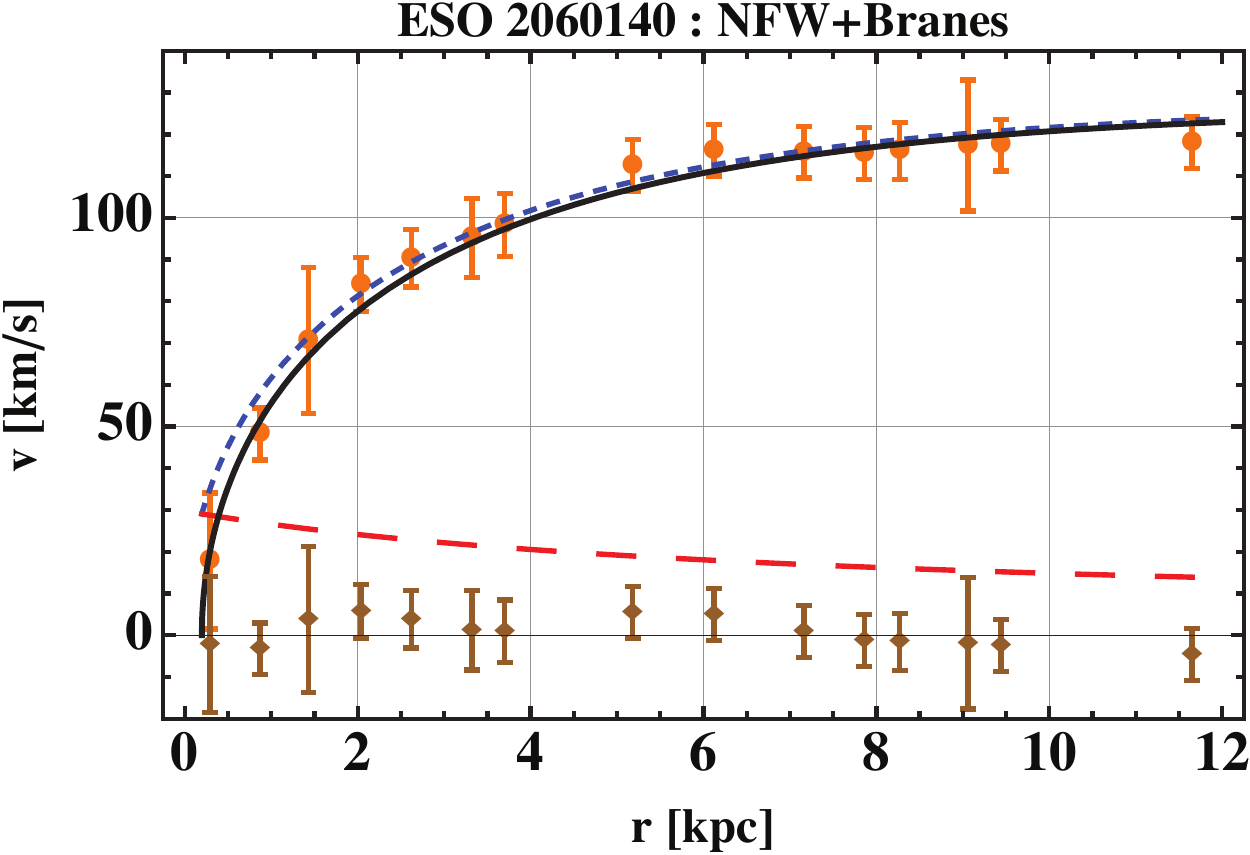}  
\includegraphics[scale=0.33]{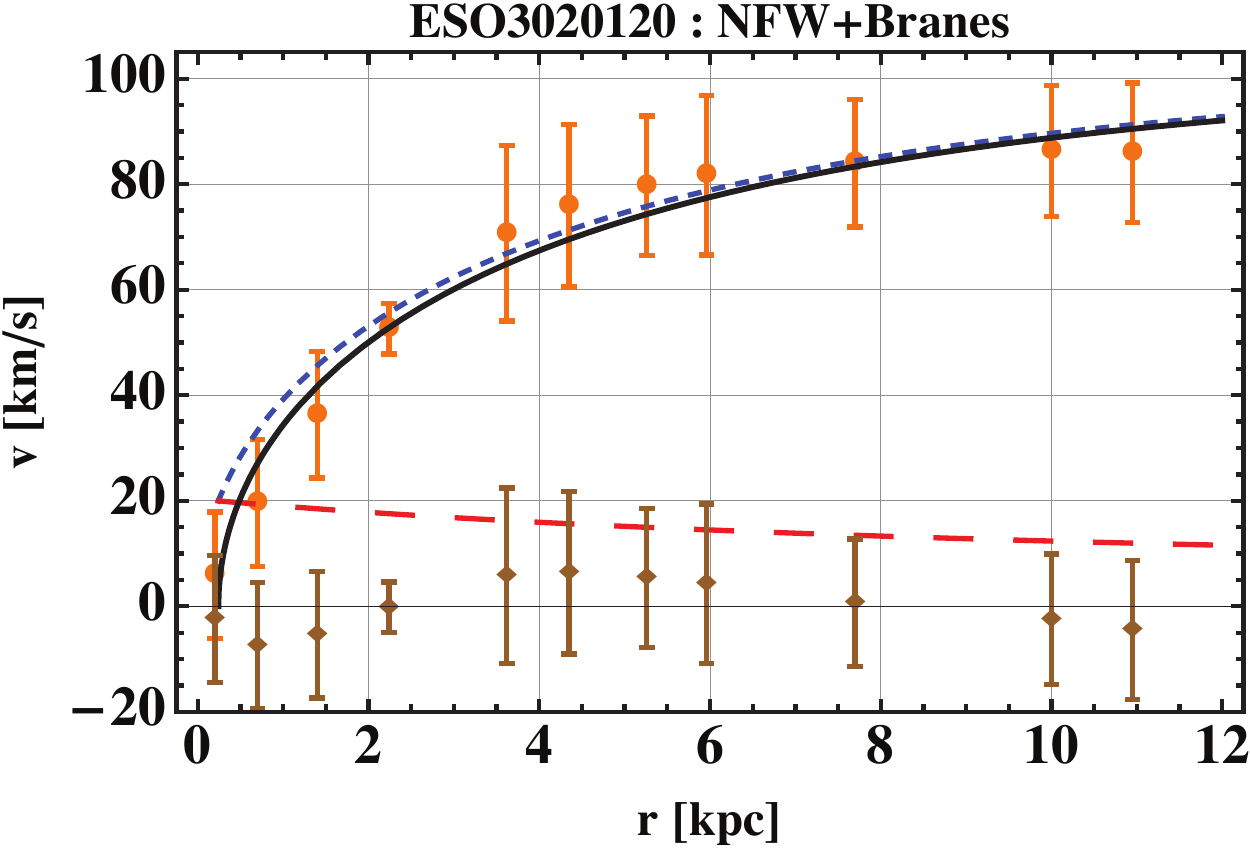} \\ 
\includegraphics[scale=0.33]{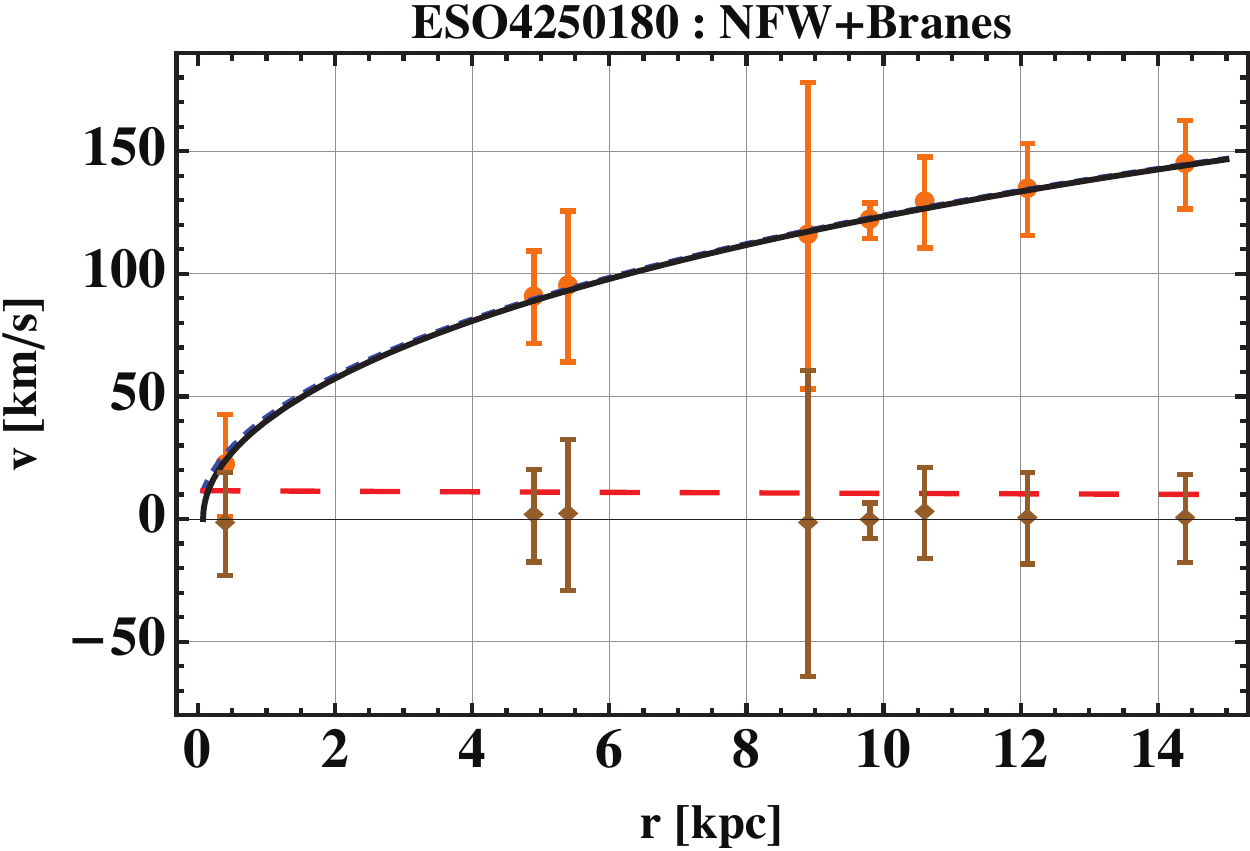} 
\includegraphics[scale=0.33]{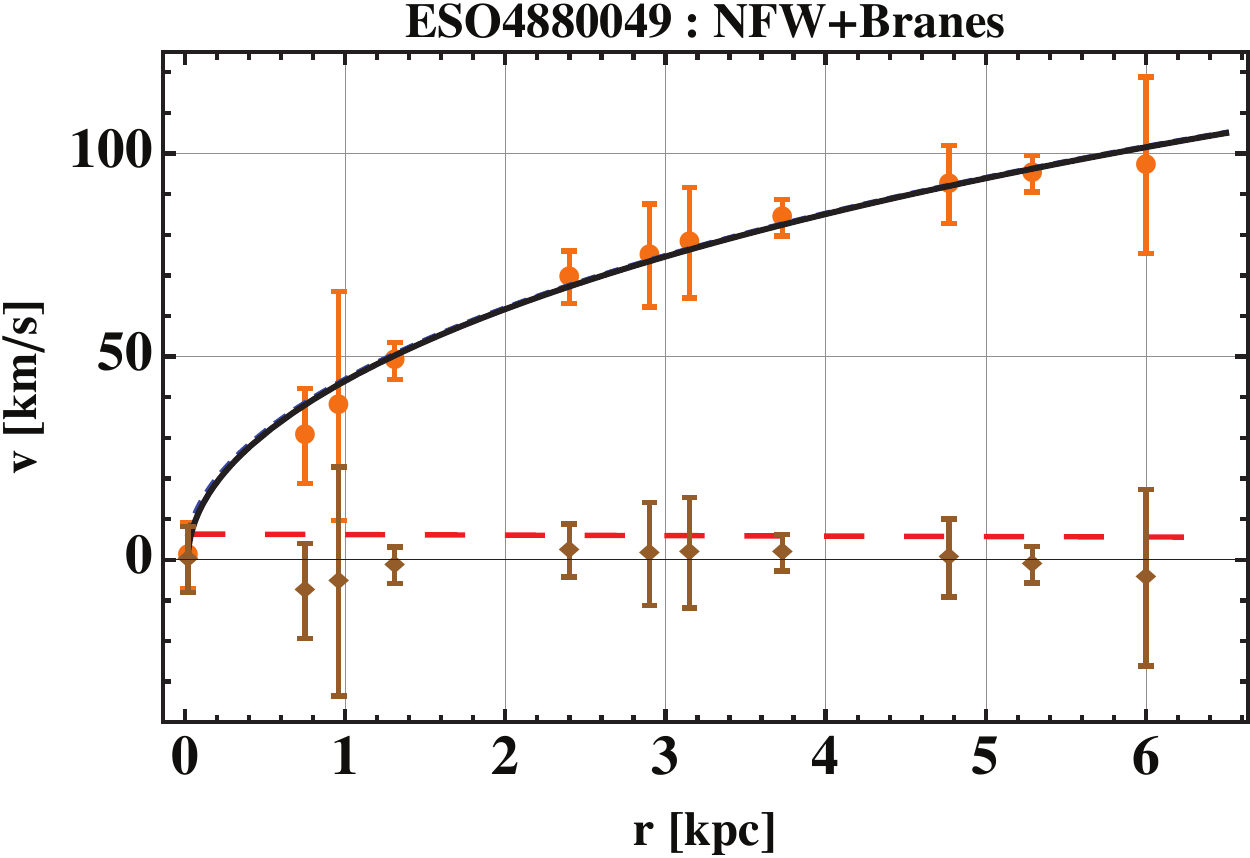} \\
\includegraphics[scale=0.33]{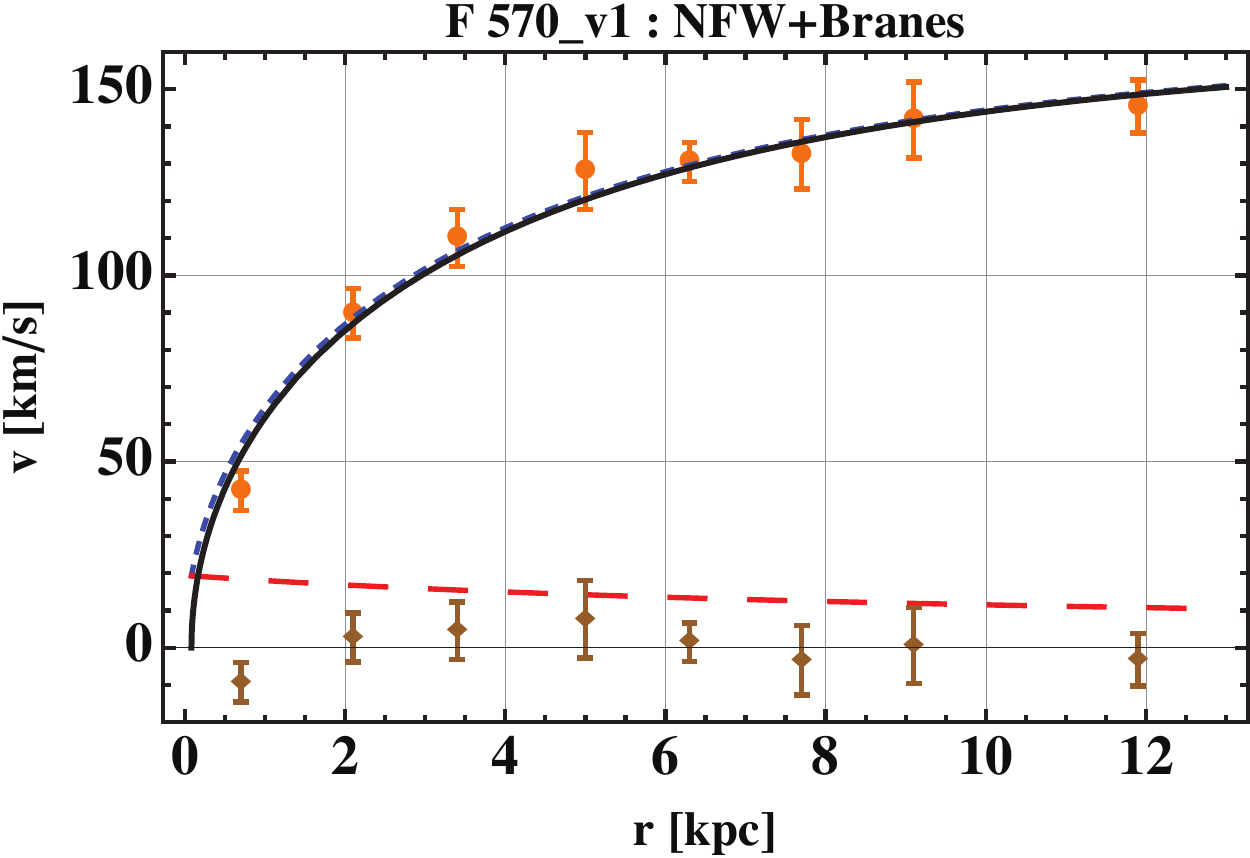} 
\includegraphics[scale=0.33]{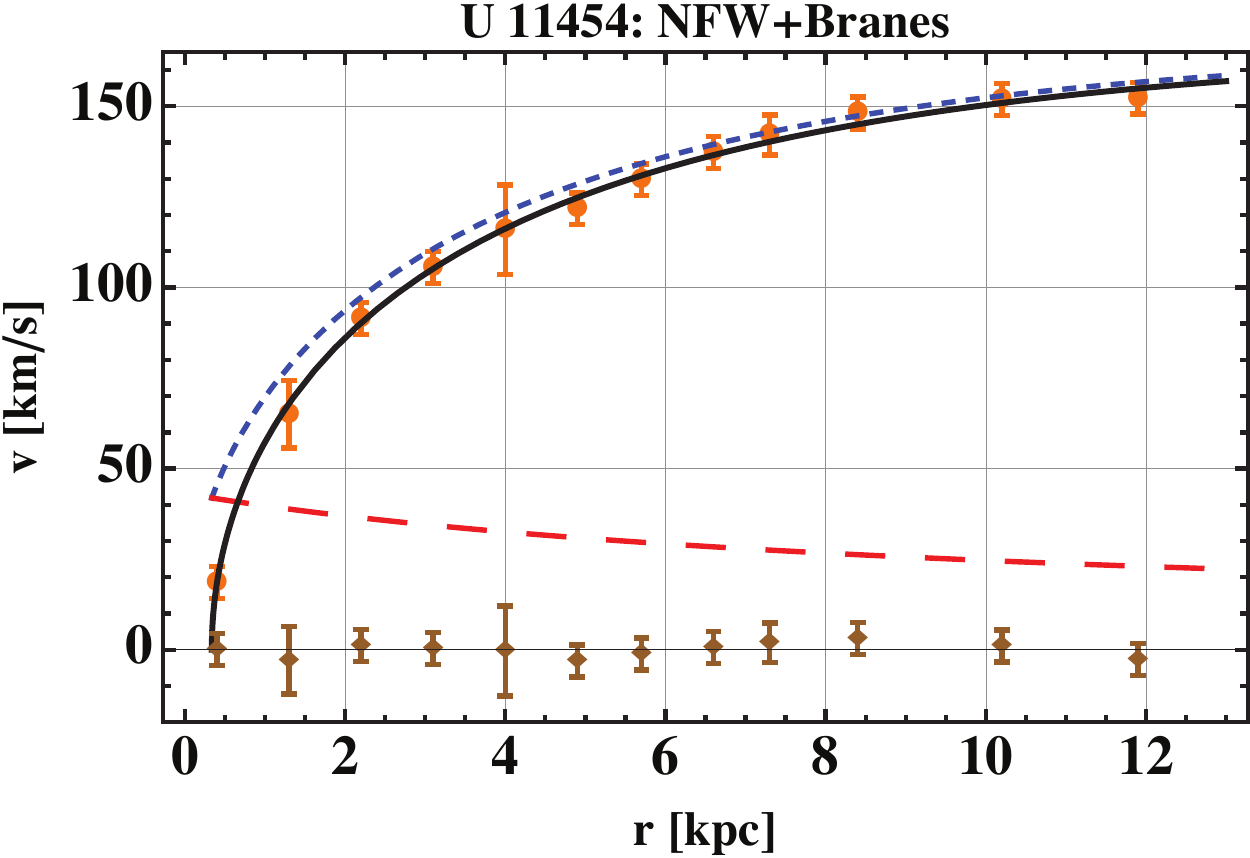} \\
\includegraphics[scale=0.33]{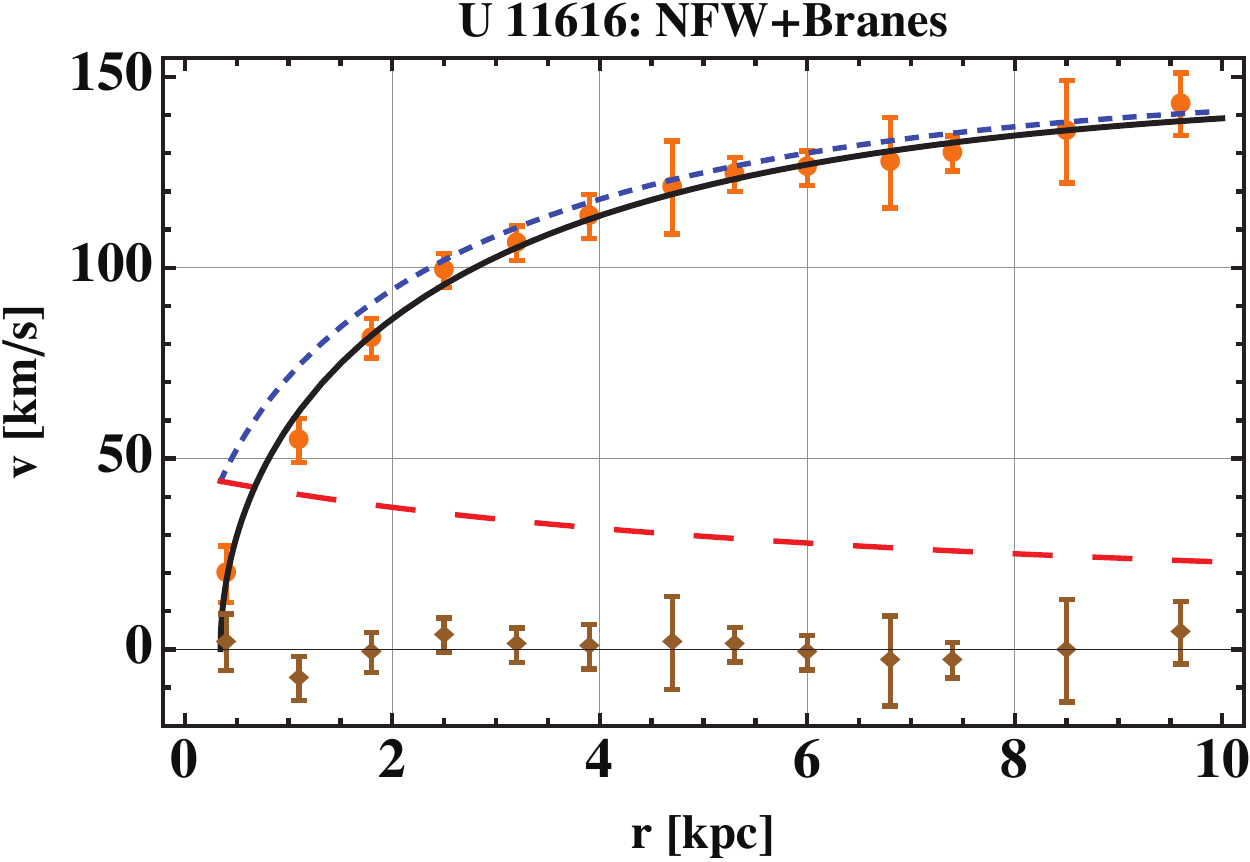} 
\includegraphics[scale=0.33]{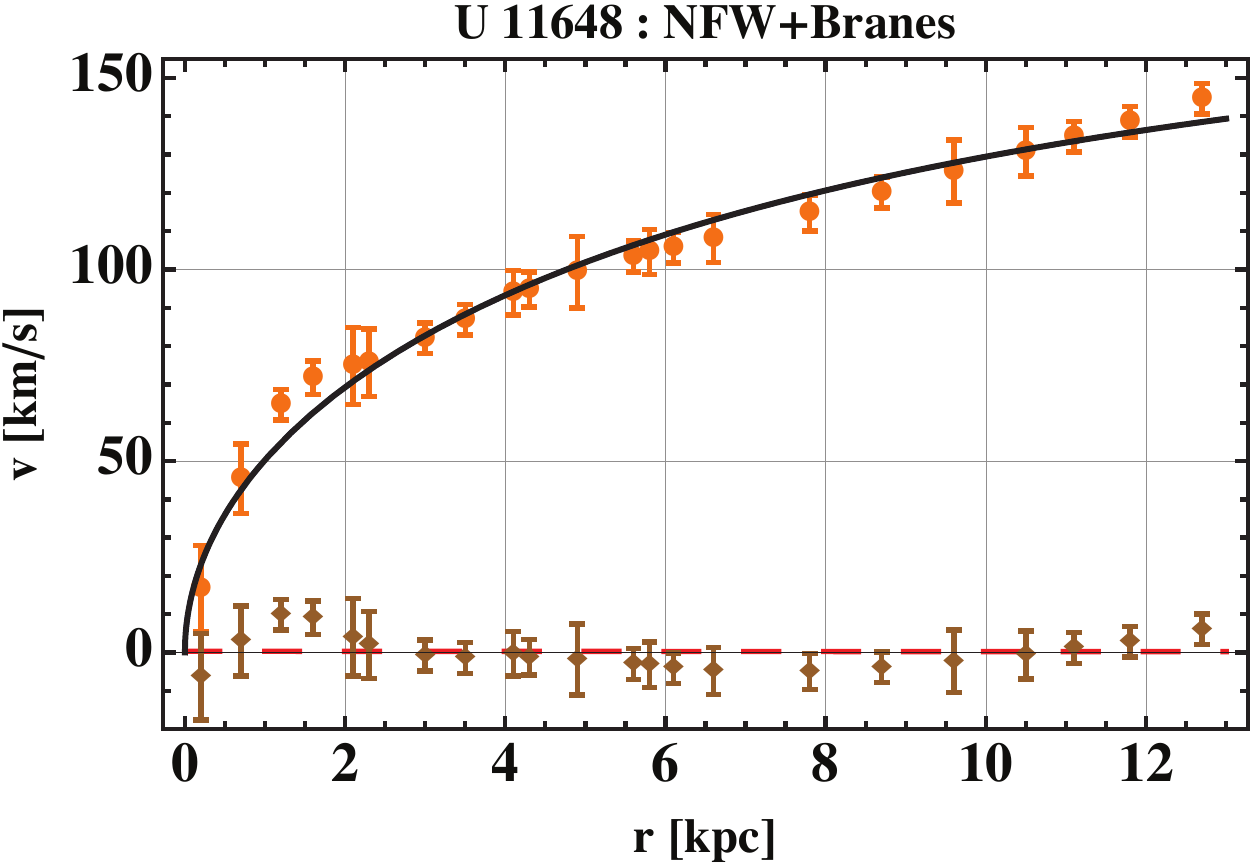} \\
\includegraphics[scale=0.33]{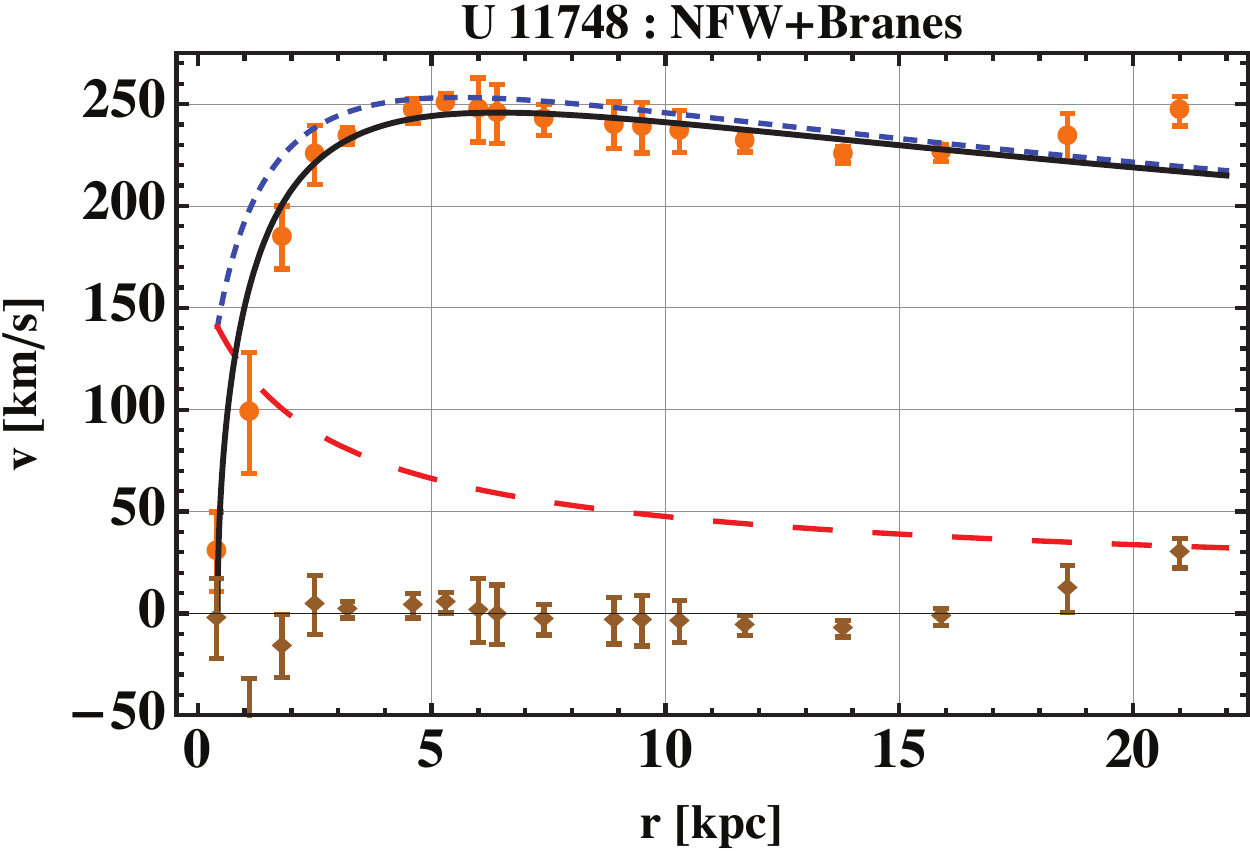} 
\includegraphics[scale=0.33]{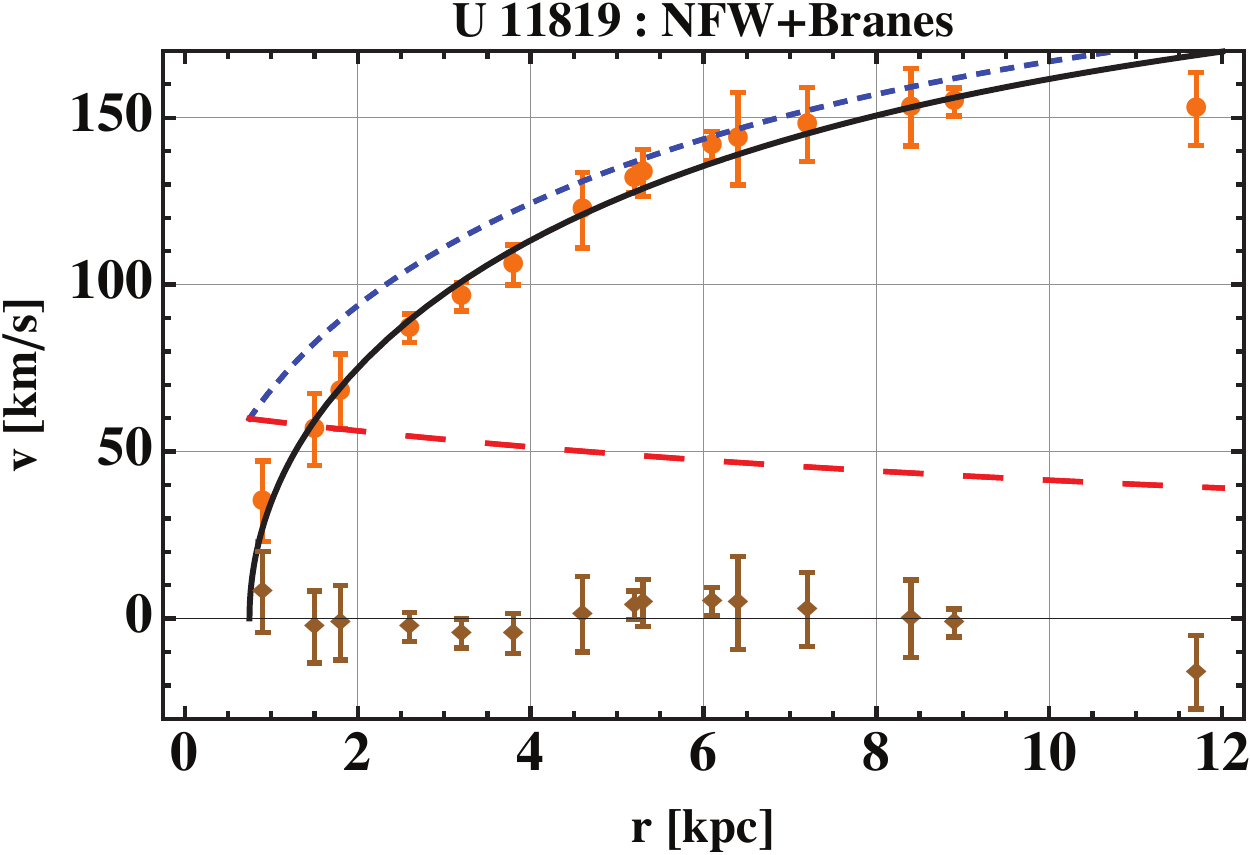} 
\caption{Group of analyzed galaxies using modified rotation velocity for NFW profile: ESO 3050090,
ESO 0140040, ESO 2060140, ESO 3020120, ESO 4250180, ESO 4880049, 570\_V1, U11454, U11616, U11648, U11748, U11819. We show in the plots: Total rotation curve (solid black line), only NFW curve (short dashed blue curve) and the rotation curve associate with the mass lost by the effect of the brane (red dashed curve).} 
\label{NFW1}
\end{figure}

In Fig.\ \ref{NFW1}, it is shown, for each galaxy in the sample of the LSB galaxies,  the theoretical fitted curve to a preferred brane tension value (solid line), 
the NFW curve and the rotation curve associated with the mass lost by the effects of branes, see Eq.\ (\ref{RCNFW}). 
In Table \ref{TableNFW} it is shown, for the sample,
the central density, central radius and $\chi_{red}^2$ values without branes; and
the central density, central radius, brane tension
and $\chi_{red}^2$ values with branes contribution.
Galaxy U 11748 is the worst fitted case with $\chi_{red}^2 = 2.163$.
For galaxies: 
ESO 4250180,
ESO 4880049,
and U 11648,
there are not clear brane effects. 
Galaxy U 11648 is an \emph{outlier} with a brane tension value of $4323.28\; M_{\odot}/\rm pc^3$ that is out of the range of preferred values of the other galaxies in the sample.
Notice that we have found a preferred range of tension values, from $0.487$ to $9.232$ $M_{\odot}/\rm pc^3$. Without the outlier, the brane tension parameter has an average value of 
$\langle\lambda\rangle_{\rm NFW}\simeq 2.51 \; M_{\odot}/\rm pc^3$ 
with a standard deviation $\sigma_{\rm NFW}\simeq 3.015 \; M_{\odot}/\rm pc^3$.

%%%%%%%%%%%%%%%%%%%%%%%%%%%
% ORIGINAL POSITION OF NFW TABLE
%%%%%%%%%%%%%%%%%%%%%%%%%%%

%%%%%%%%%%%%%%%%%%%%%%%%%%%%%%%%%%%%%%%%%%%%%%%
\subsection{Results: Burkert+Branes profile}
%%%%%%%%%%%%%%%%%%%%%%%%%%%%%%%%%%%%%%%%%%%%%%%

In the case of Burkert DM density profile, we have also estimated the parameters of the Burkert+branes model 
and were compared with Burkert model without branes, minimizing the appropriate $\chi^2_{red}$, Eq.\ (\ref{chi2Eq}) with Eq.\ (\ref{RCBurkert}), for the sample of observed rotation curves. We have considered that $\lambda > \rho_s$ must be fulfilled.

\begin{figure}
\includegraphics[scale=0.33]{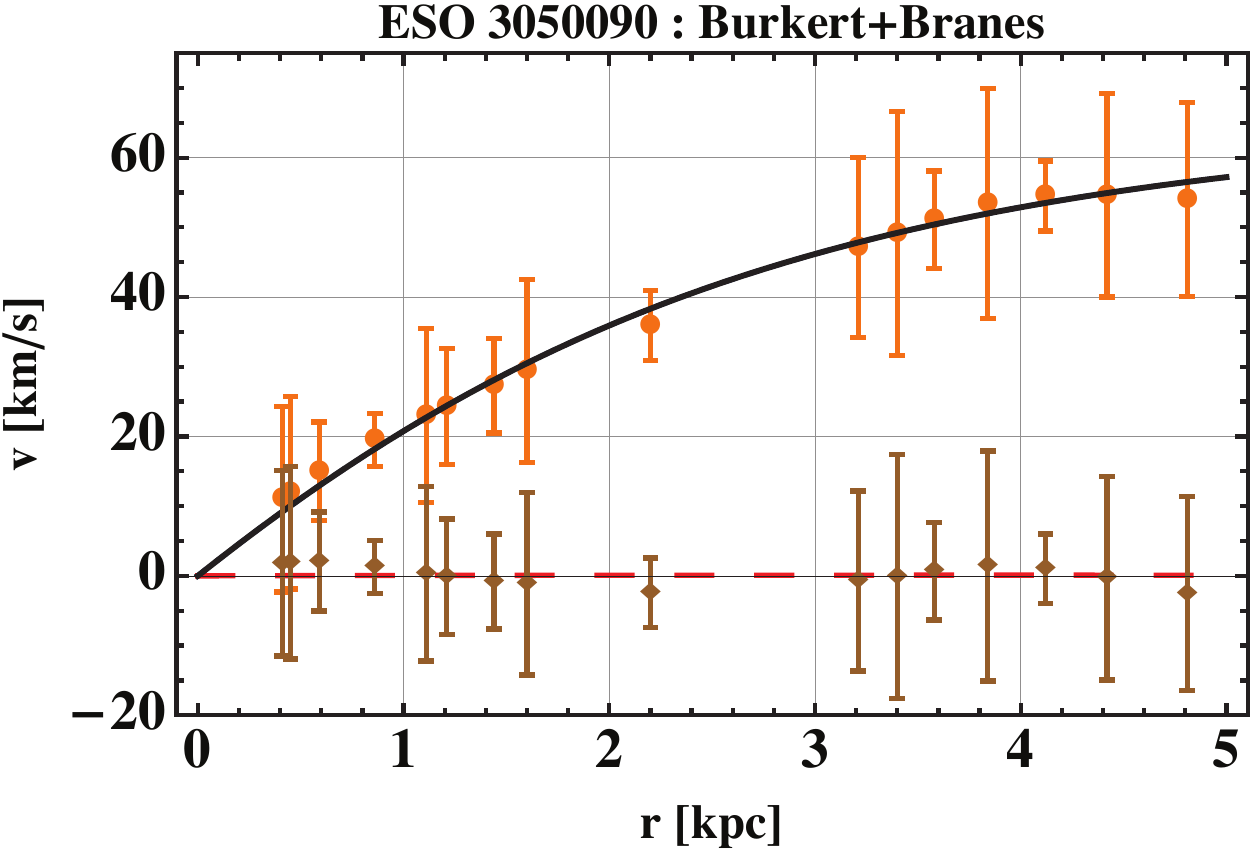} 
\includegraphics[scale=0.33]{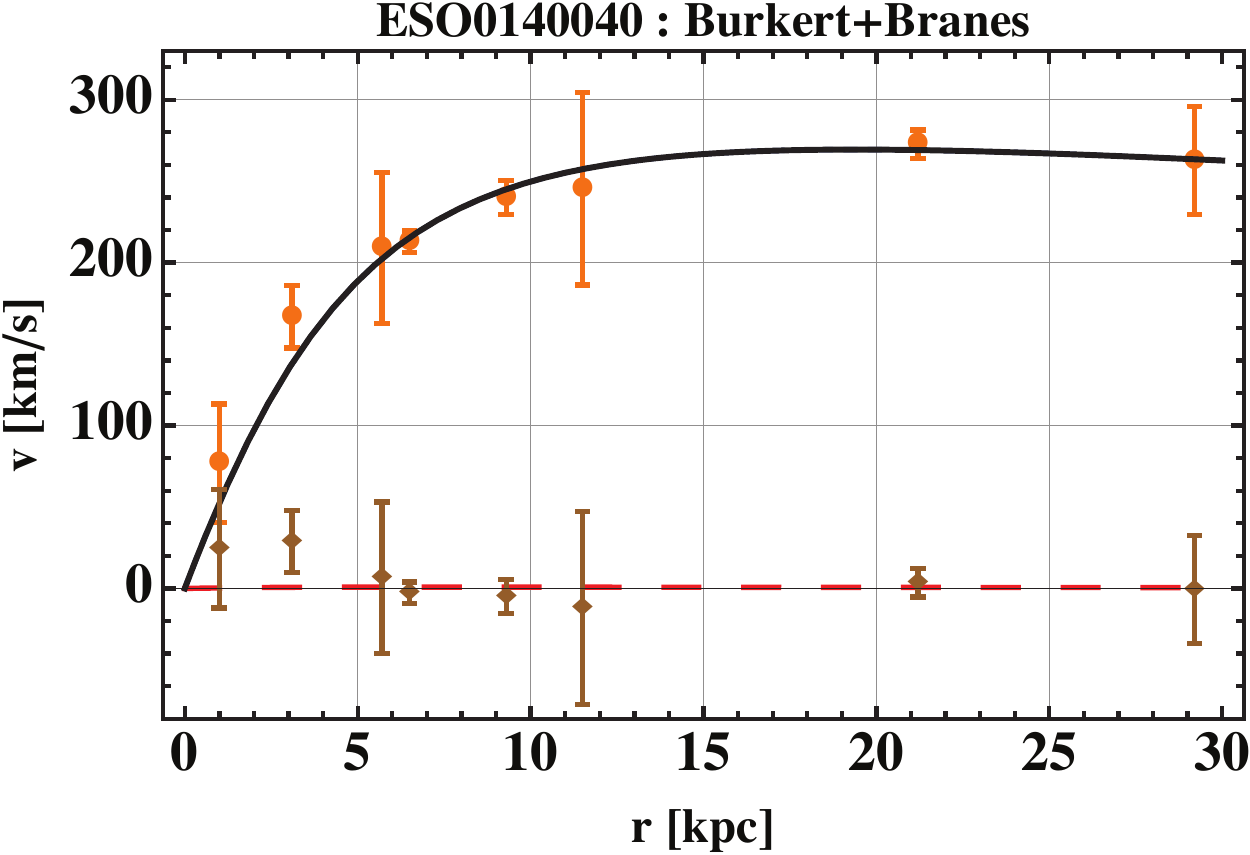} \\
\includegraphics[scale=0.33]{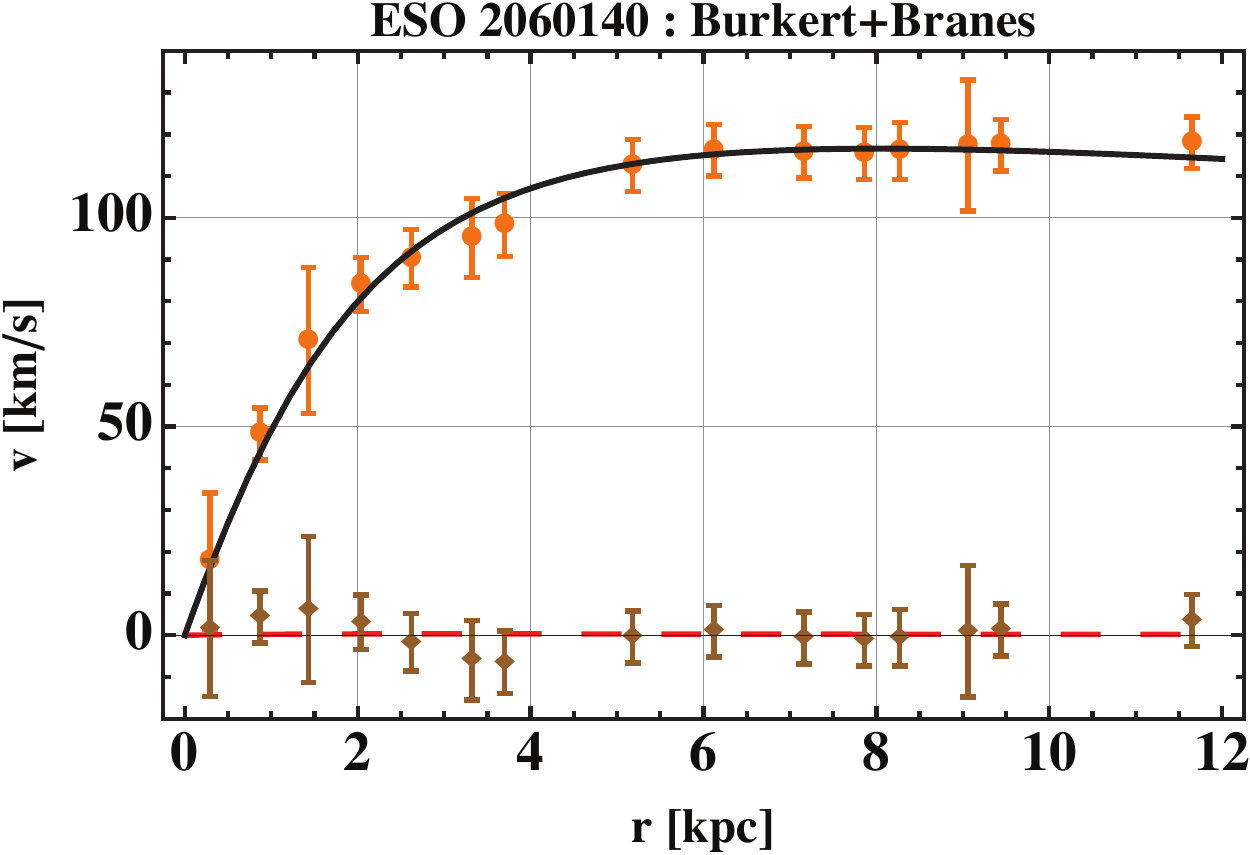} 
\includegraphics[scale=0.33]{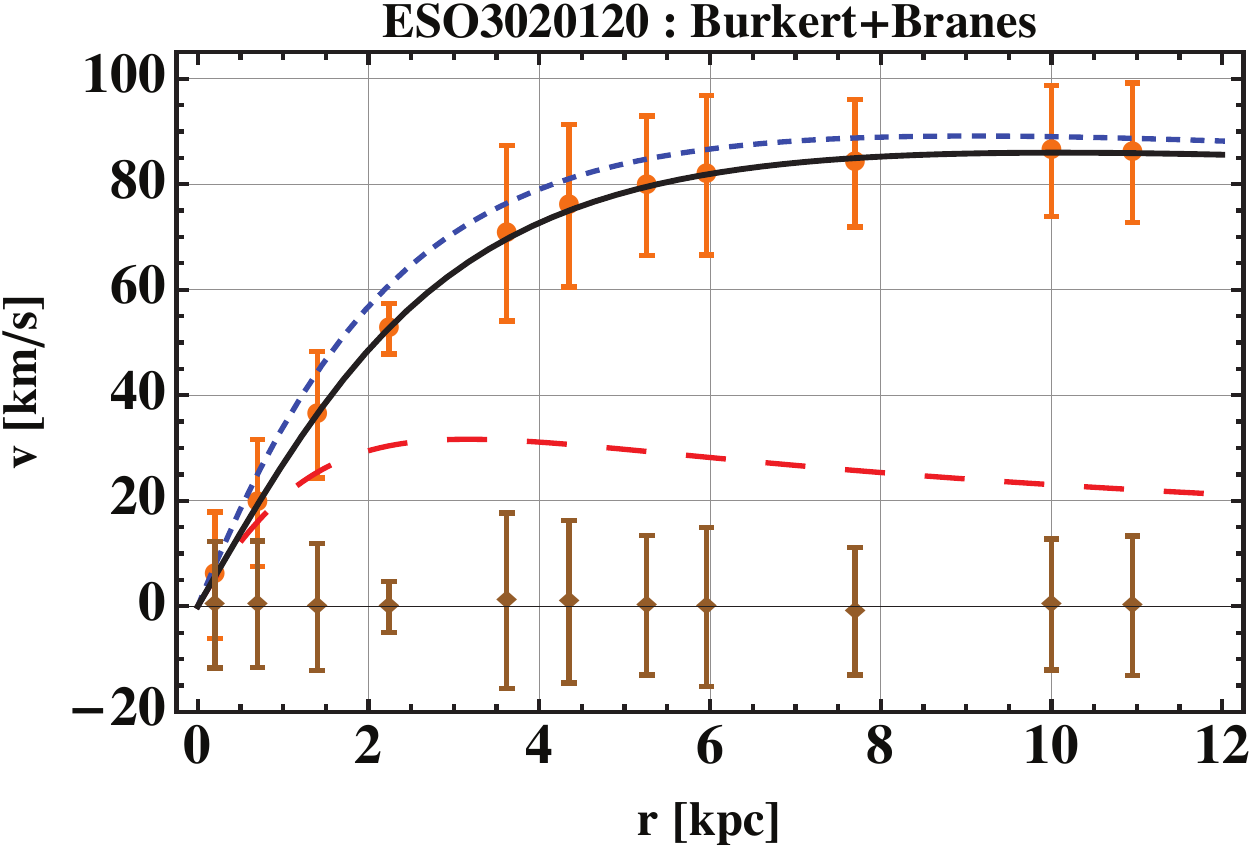} \\ 
\includegraphics[scale=0.33]{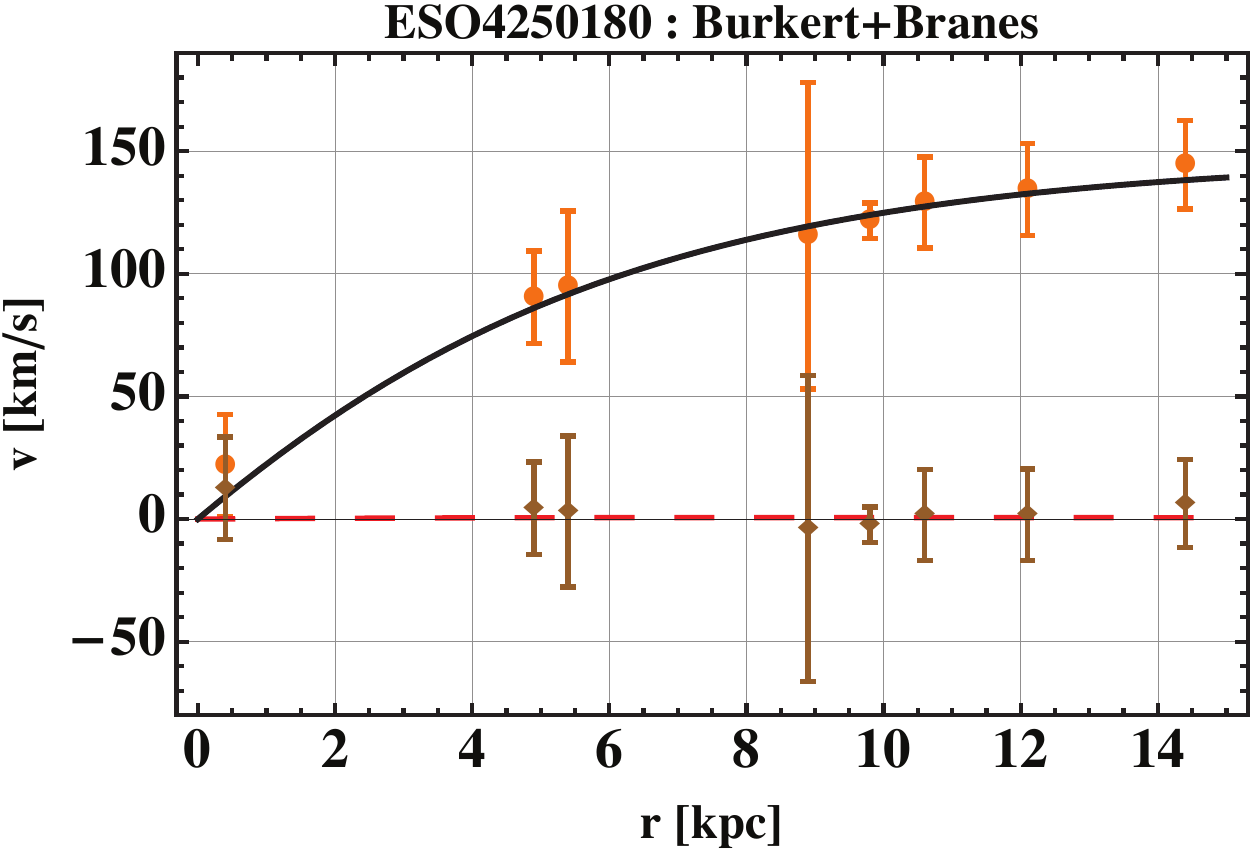} 
\includegraphics[scale=0.33]{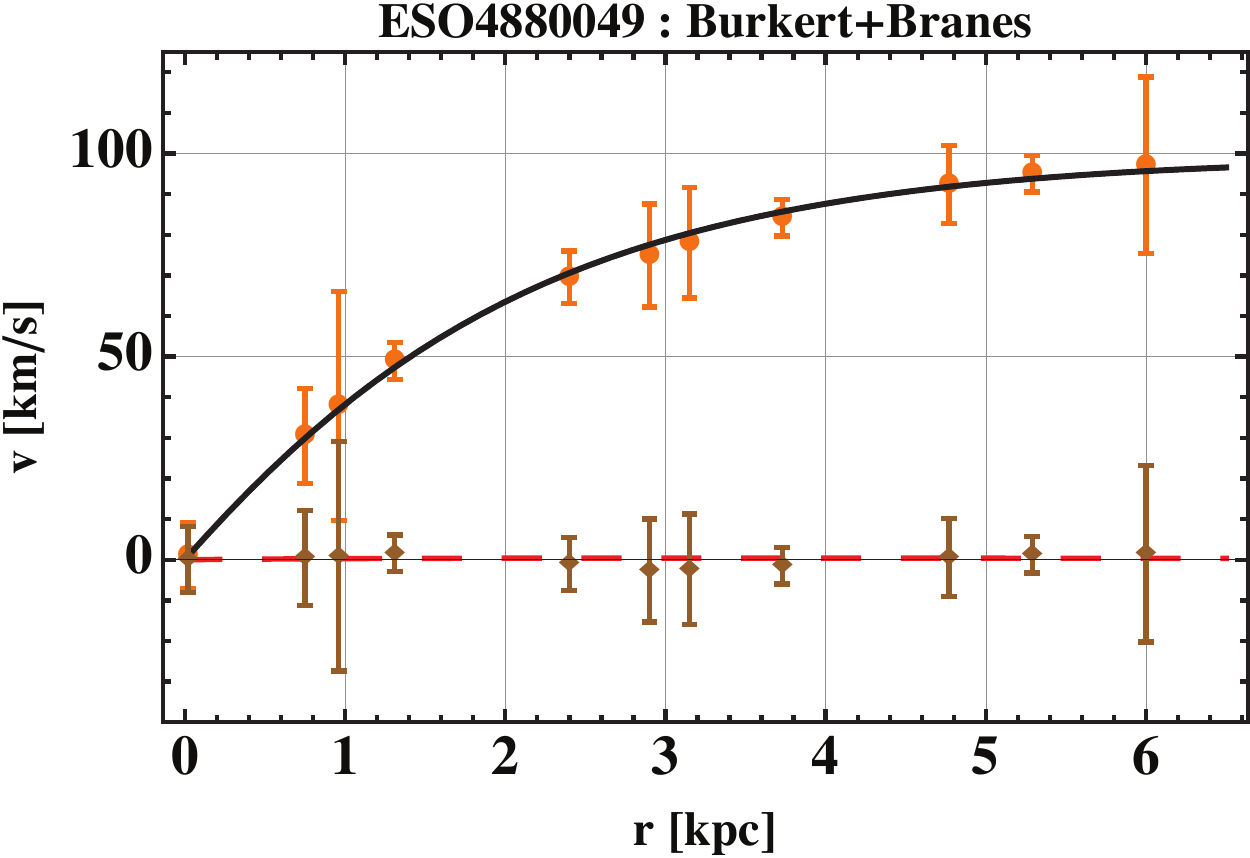} \\
\includegraphics[scale=0.33]{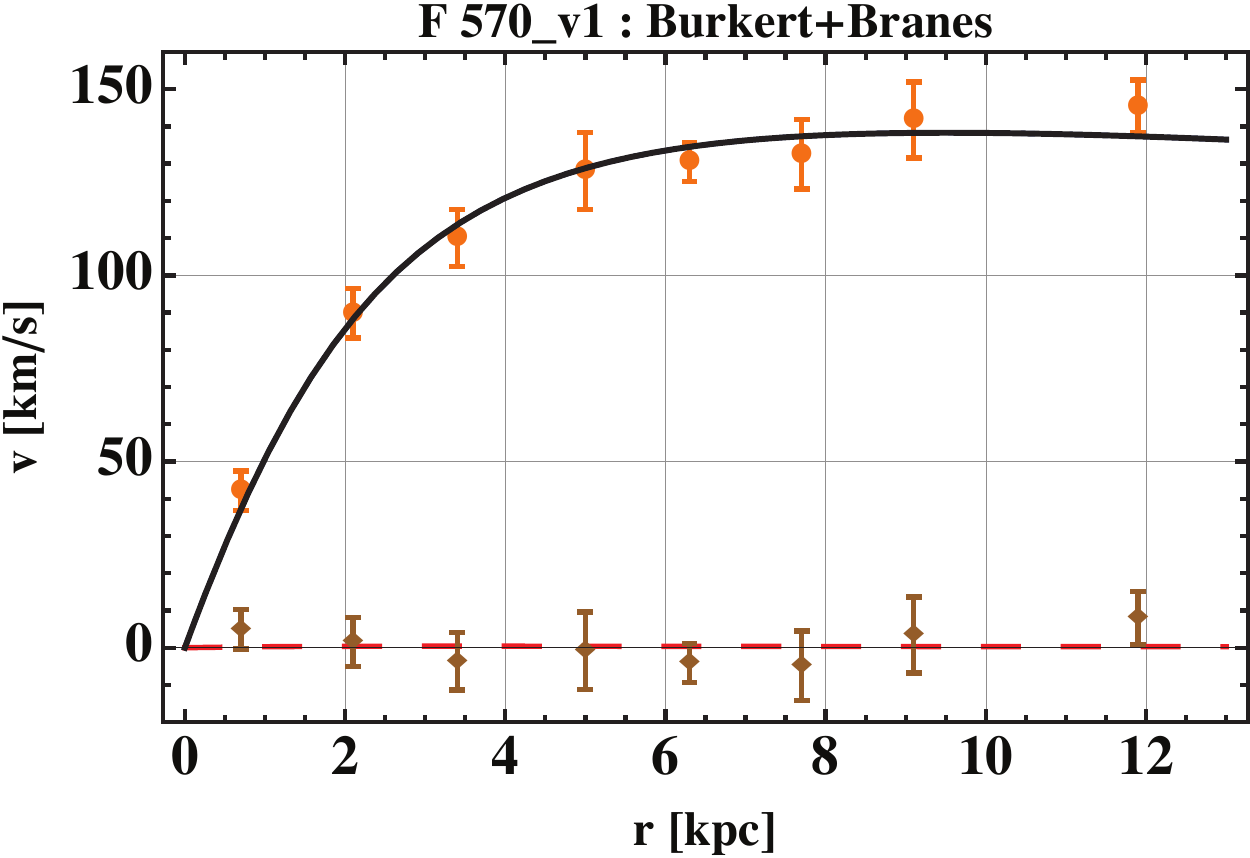} 
\includegraphics[scale=0.33]{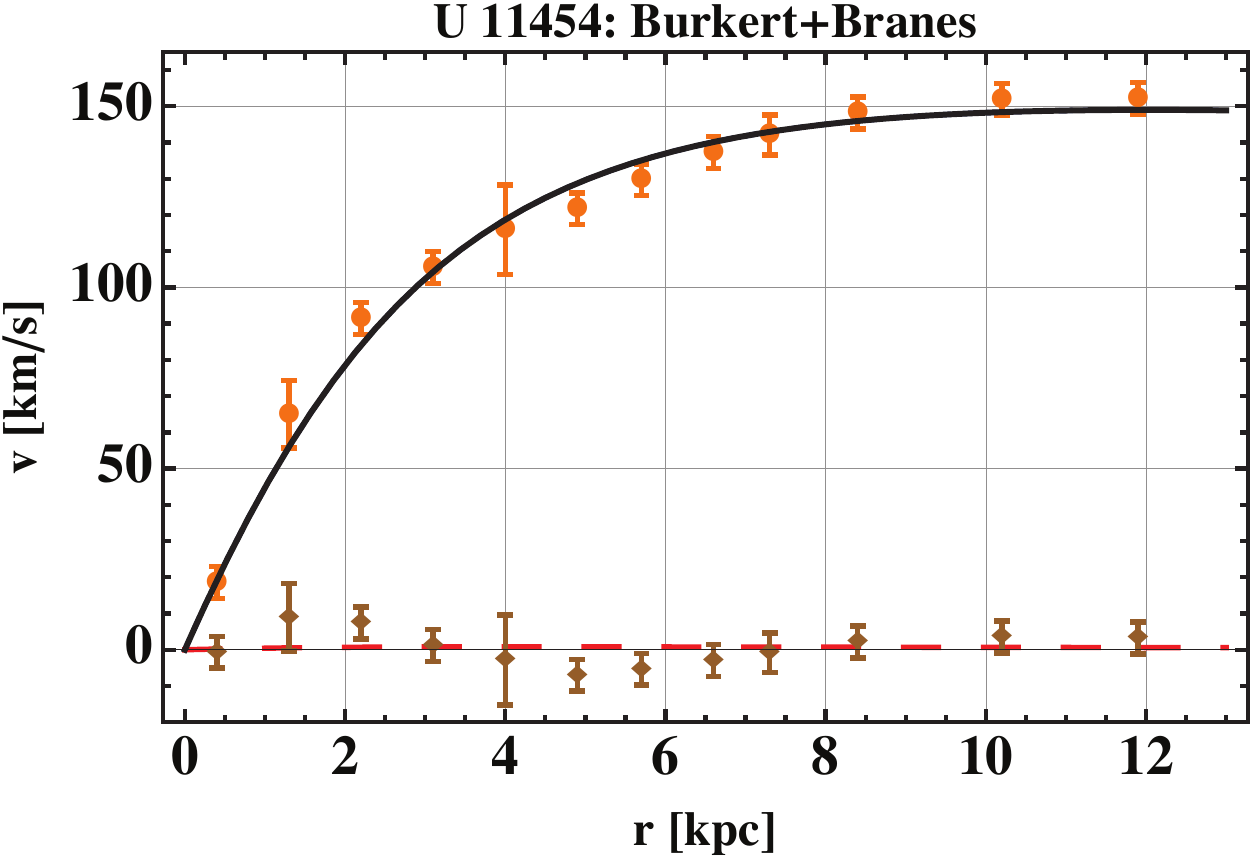} \\
\includegraphics[scale=0.33]{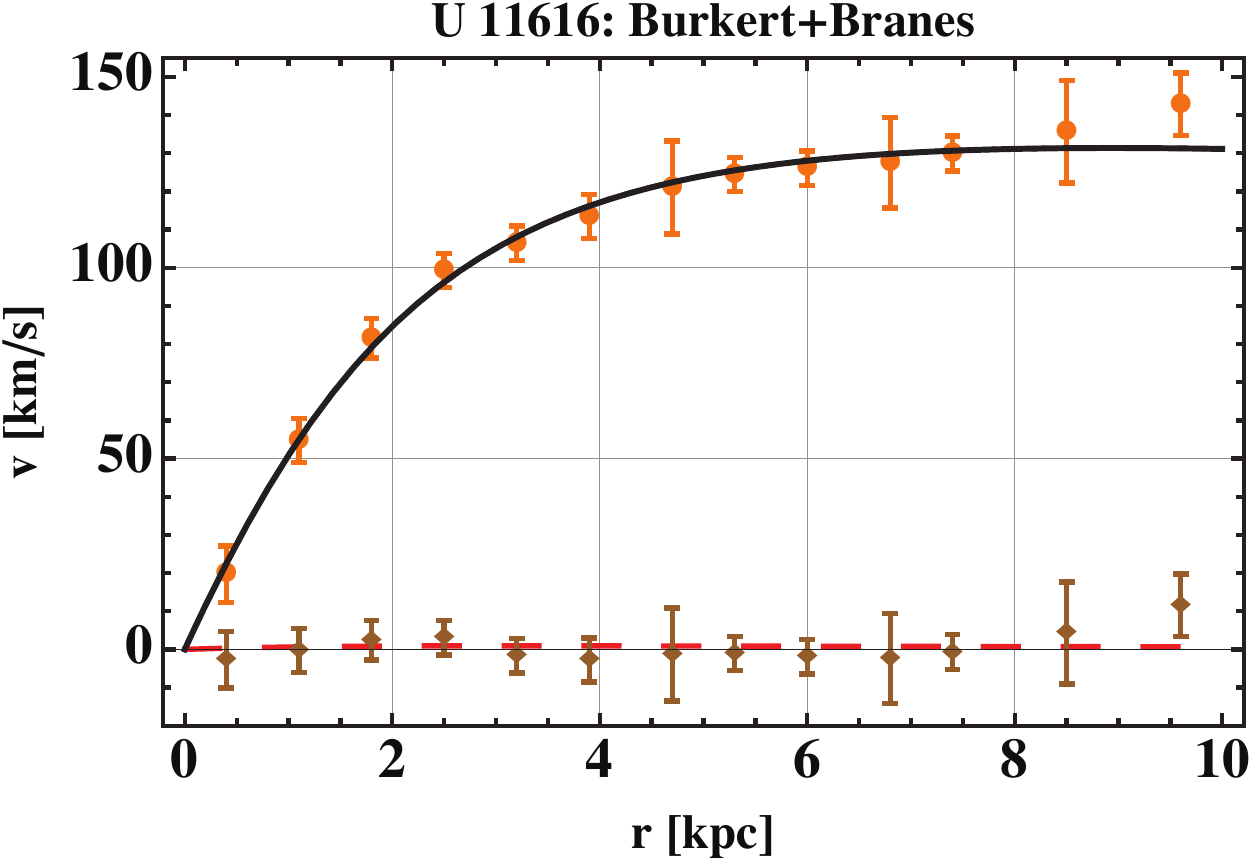} 
\includegraphics[scale=0.33]{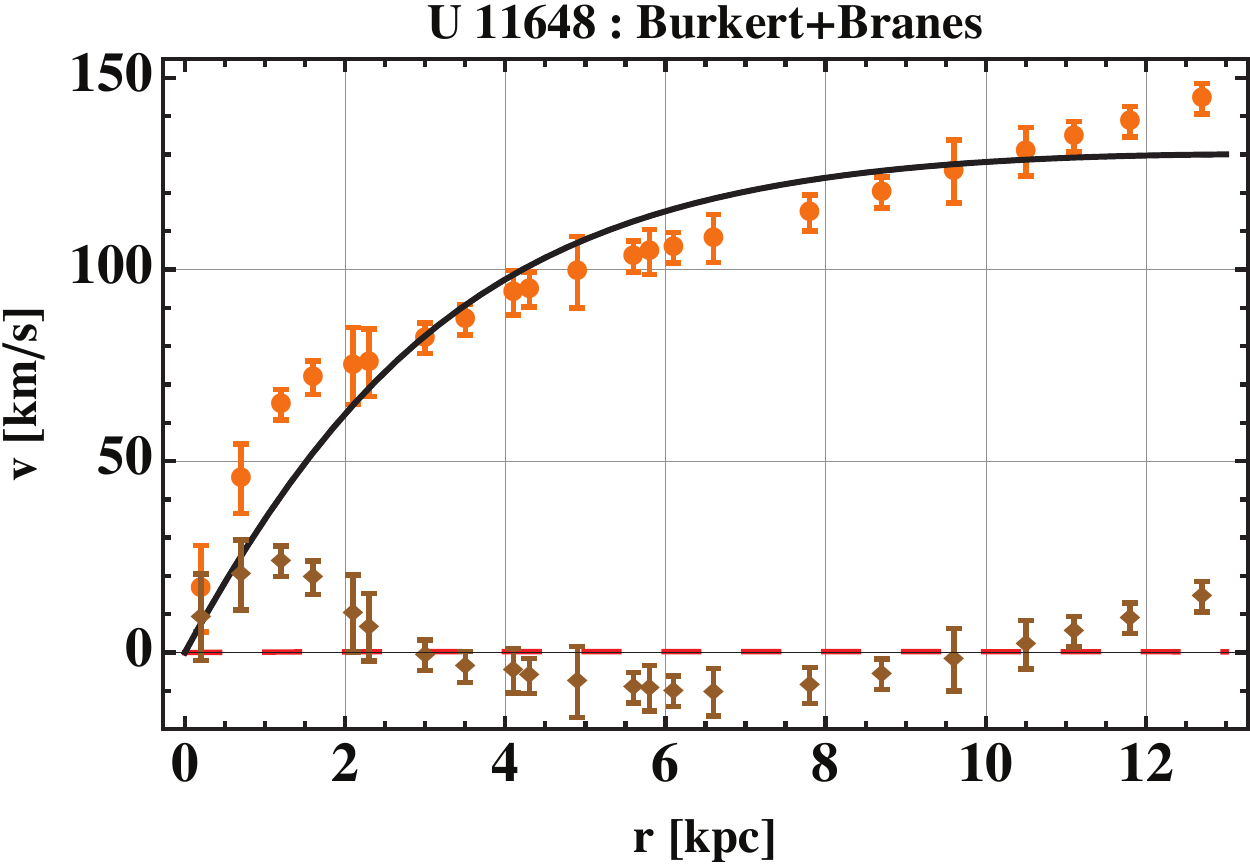} \\
\includegraphics[scale=0.33]{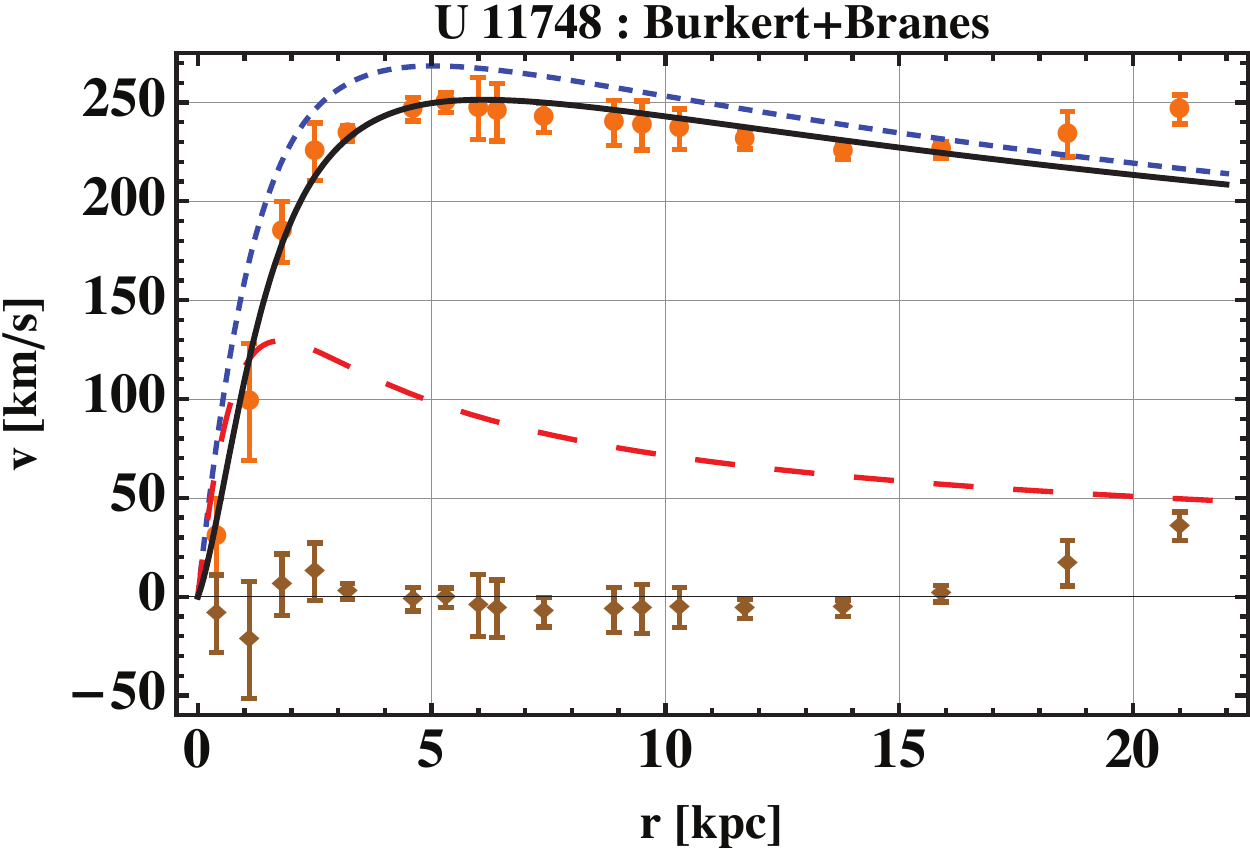} 
\includegraphics[scale=0.33]{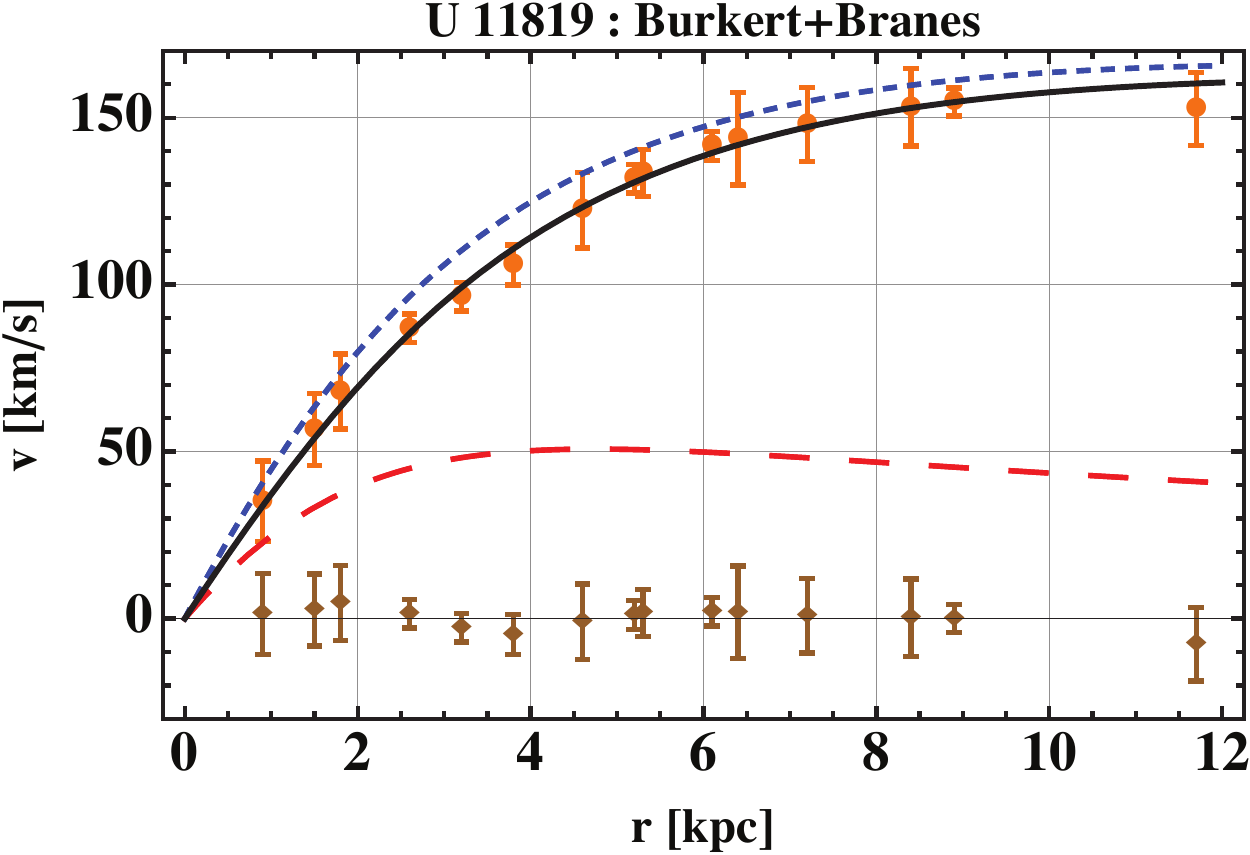}
\caption{Group of analyzed galaxies using modified rotation velocity for Burkert profile: ESO 3050090,
ESO 0140040, ESO 2060140, ESO 3020120, ESO 4250180, ESO 4880049, 570\_V1, U11454, U11616, U11648, U11748, U11819. We show in the plots: Total rotation curve (solid black line), only Burkert curve (short dashed blue curve) and the rotation curve associate with the mass lost by the effect of the brane (red dashed curve).} 
\label{Burkert1}
\end{figure}

The results are shown in Fig.\ \ref{Burkert1}, where it is plotted the fit to a preferred brane tension value, remarking the total rotation curve (solid line), the Burkert DM density contribution curve (blue short-dashed line) and the rotation curve associated with the mass lost by the effects of branes (red dashed line), see Eq.\ (\ref{RCBurkert}). 
In Table \ref{TableBurkert} it is shown the fitted values for the central density, central radius and the corresponding value of the $\chi_{red}^2$ without brane contribution; and the fitted values for
the central density, central radius, brane tension, and theirs $\chi_{red}^2$ values with brane contribution.
The worst fitted (high values of $\chi_{red}^2$) galaxies are: U 11648 and U 11748.
Galaxies ESO 3020120, U 11748, and
U 11819 show a clear brane effects and also are outliers. The main tendency is that $\lambda$ has values of the order of $10^3 \;M_{\odot}/\rm pc^3$ or above, approximately.
The brane tension parameter, without the outliers, for the DM Burkert profile case has an average value of 
$\langle\lambda\rangle_{\rm Burk}\simeq 3192.02 \;M_{\odot}/\rm pc^3$, 
and a standard deviation of 
$\sigma_{\rm Burk}\simeq 2174.97 \; M_{\odot}/\rm pc^3$.

%%%%%%%%%%%%%%%%%%%%%%%%%%%
% ORIGINAL POSITION OF BURKERT TABLE
%%%%%%%%%%%%%%%%%%%%%%%%%%%

%%%%%%%%%%%%%%%%%%%%%%%%%%%%%%%%%%%%%%%%%%%%%%%
\subsection{Results: a synthetic rotation curve}
%%%%%%%%%%%%%%%%%%%%%%%%%%%%%%%%%%%%%%%%%%%%%%%

Finally, we show the fitting results of the DM models plus brane's contribution to a synthetic rotation curve. This synthetic rotation curve was made of 40 rotation curves of galaxies with magnitudes around $M_I = -18.5$\cite{Salucci1}.
These 40 rotation curves came out of 1100 galaxies that gave the universal rotation curve for spirals. For this sample of low luminosity galaxies, of $M_I = -18.5$, it was shown that the baryonic disk has a very small contribution (for details see reference\cite{Salucci1}).

In this subsection we are now using units such $G=R_{opt}=V(R_{opt})=1$, where $R_{opt}$ and $V(R_{opt})$ are the optical radius and the velocity at the optical radius, respectively. $R_{opt}$ is the radius encompassing 83 per cent of the total integrated light. For an exponential disk with a surface brightness given by: $I(r) \propto \exp(-r/R_D)$, we have that $R_{opt}=3.2 R_D$\cite{Salucci1}.

\begin{figure}
\includegraphics[scale=0.33]{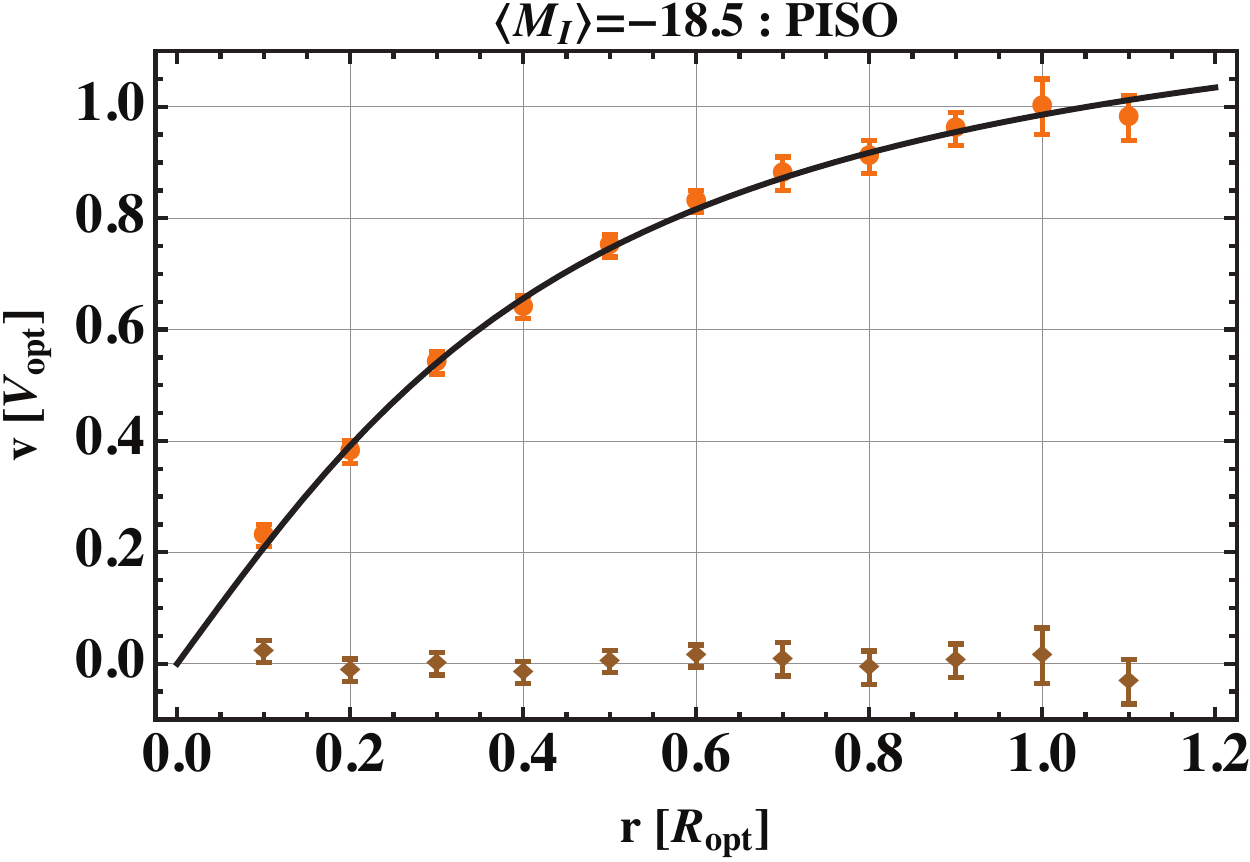} 
\includegraphics[scale=0.33]{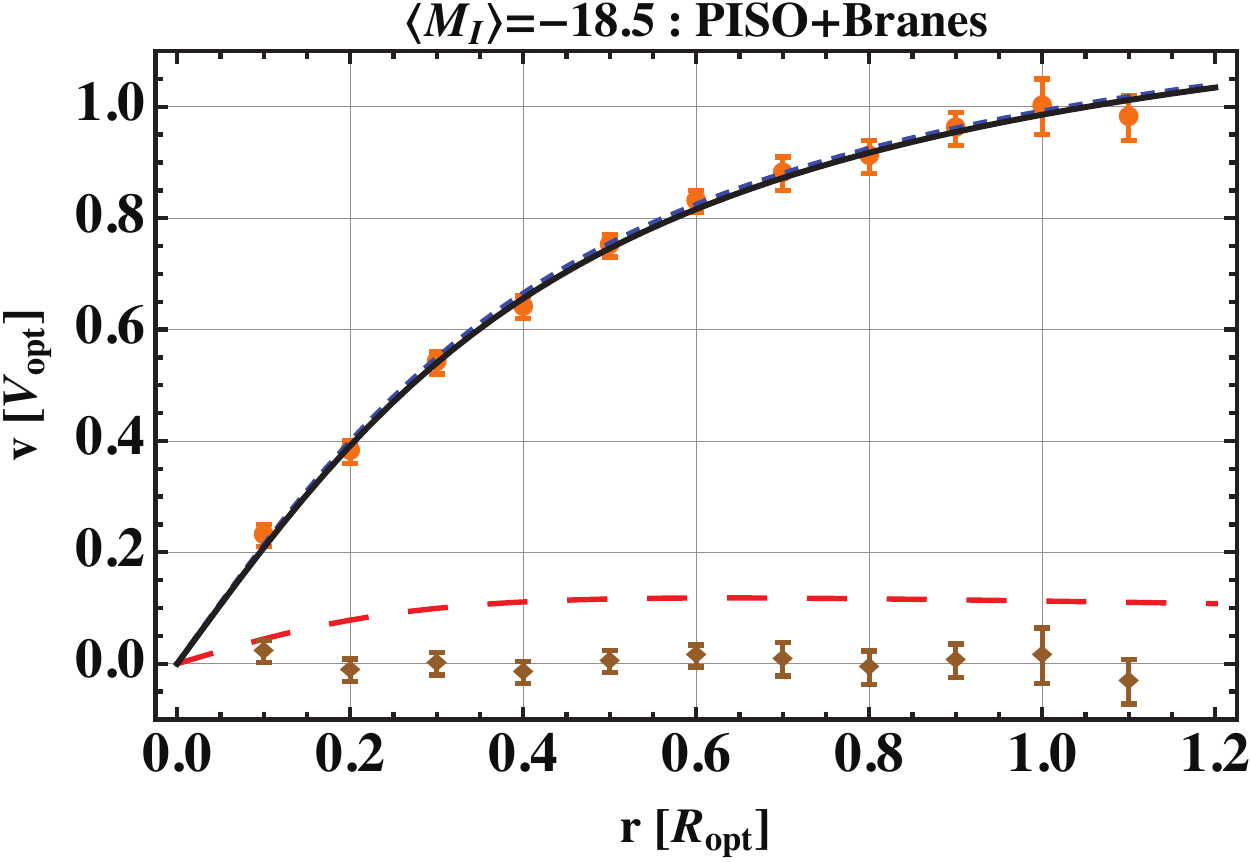} 
\includegraphics[scale=0.33]{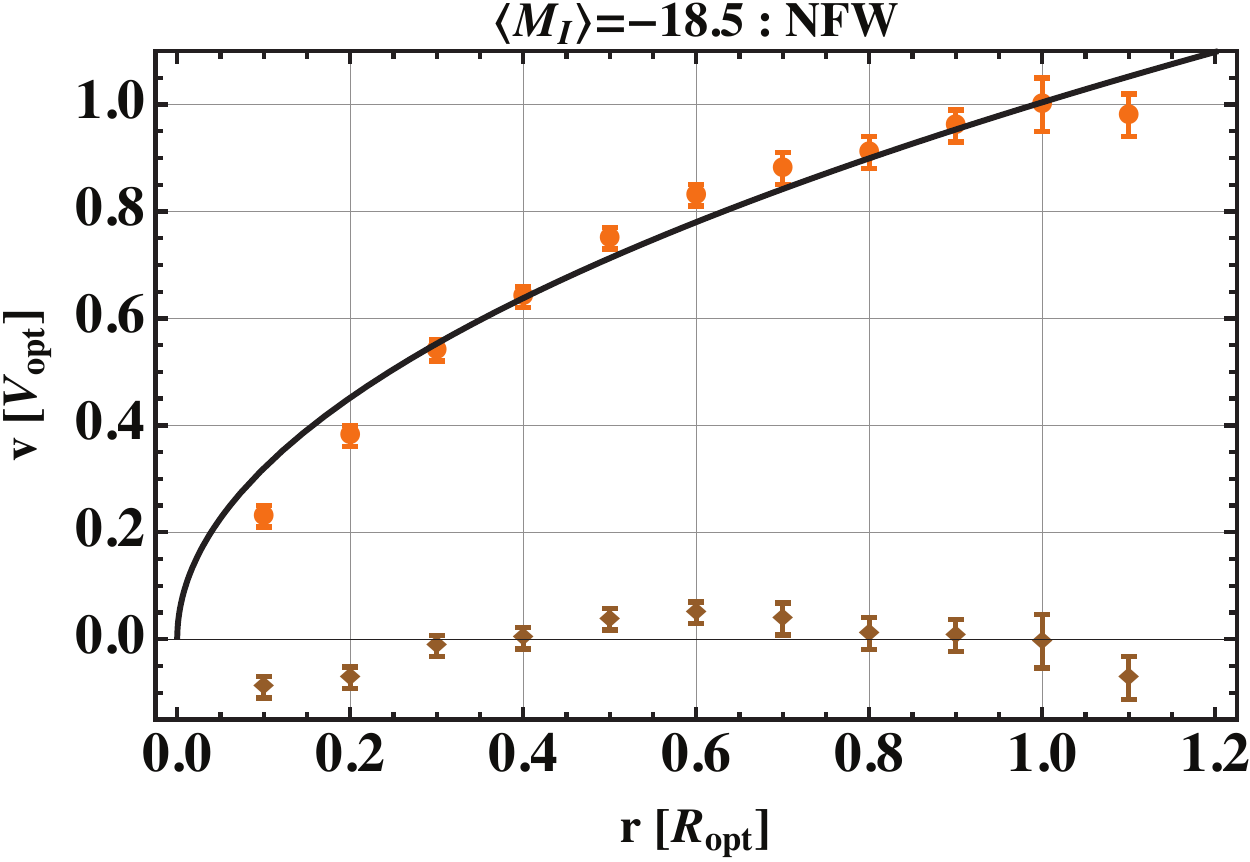} 
\includegraphics[scale=0.33]{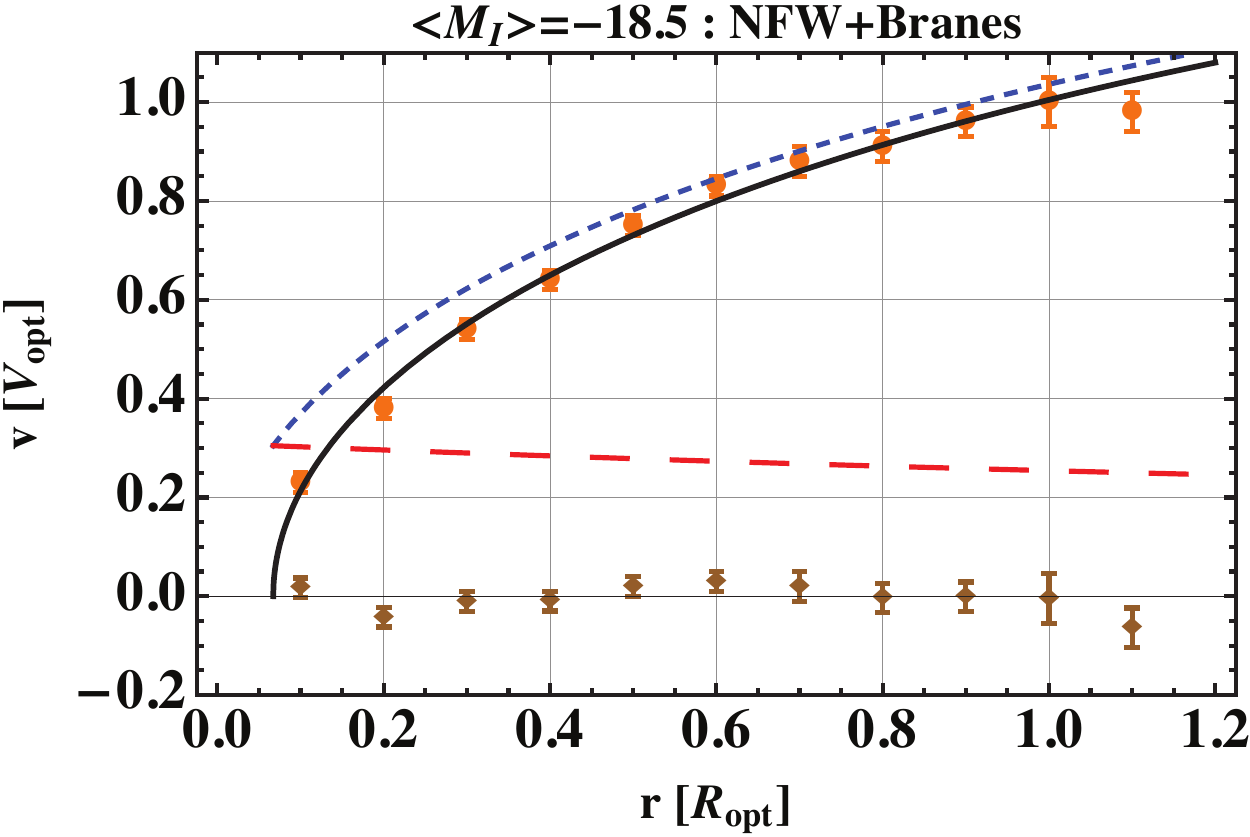} 
\includegraphics[scale=0.33]{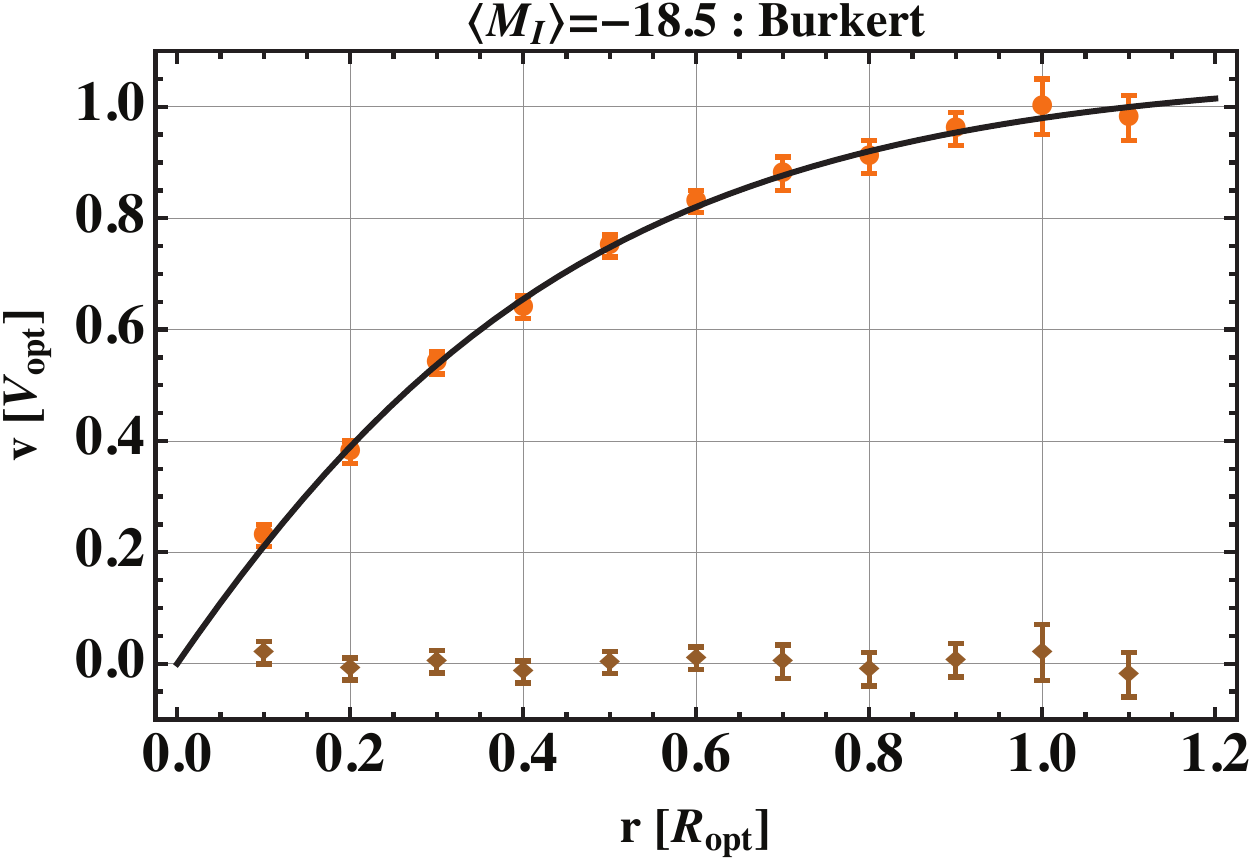} 
\includegraphics[scale=0.33]{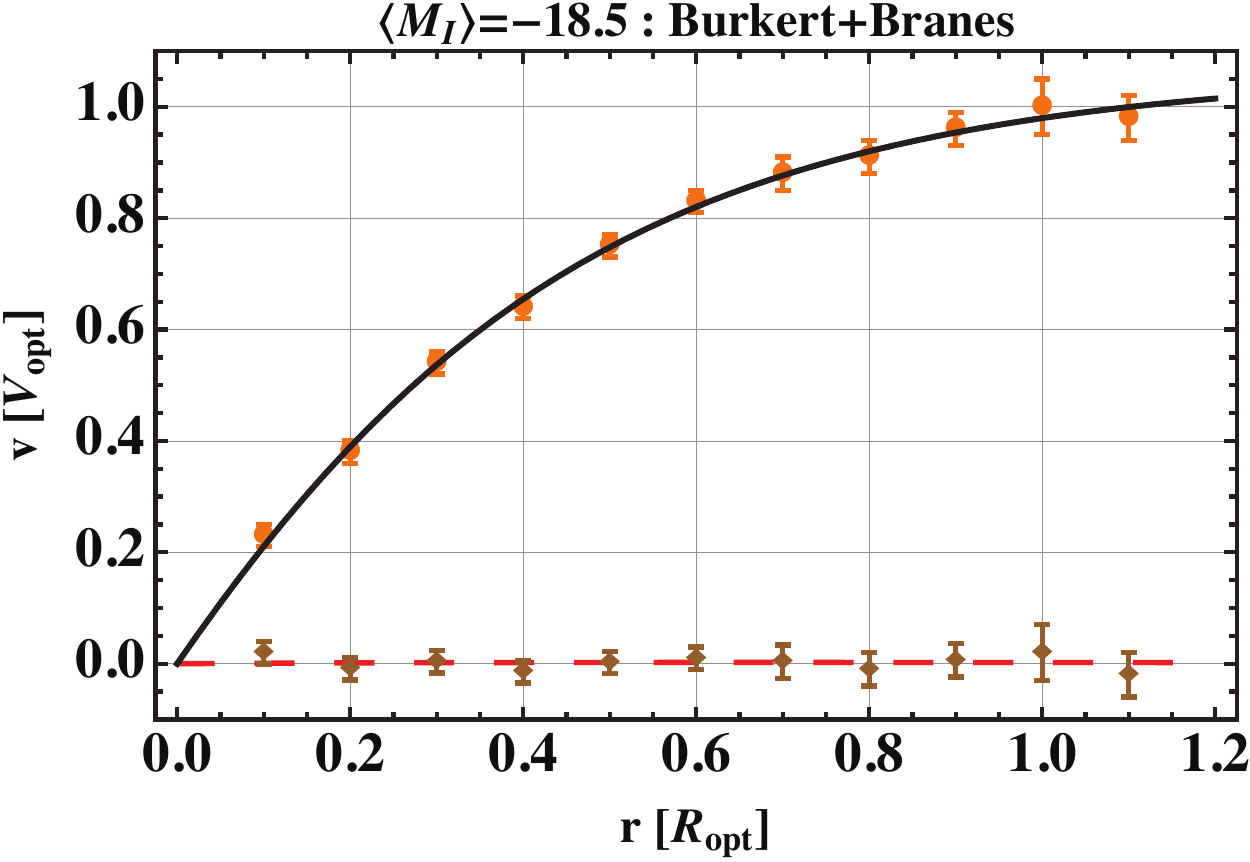} 
\caption{Synthetic rotation curve of galaxies with magnitud $M_I=-18.5$.
Left panels: rotation curves fitted without branes. Right panels: rotation curves fitted with branes.
First row is for PISO model; second row is for NFW model and third row is for Burkert model. 
We show in the plots: Total rotation curve (solid black line), only DM model curve (short dashed blue curve) and the rotation curve associate with the mass lost by the effect of the brane (red dashed curve).} 
\label{SM185}
\end{figure}

In figure \ref{SM185} we show the synthetic rotation curve and the fitting results using PISO, NFW and Burkert profiles with and without brane's contribution.
As we can see in Table \ref{TableSynthetic} the same trend is observed in the brane's tension values as compared with the results for the LSB catalog analyzed above using PISO, NFW and Burkert as a DM profiles: lower value is obtained for NFW model and higher values is obtained for Burkert density profile.

Given that this synthetic rotation curve is built from 40 rotation curves of real spirals, the values of the brane's tension in table \ref{TableSynthetic} is representative of all these rotation curves. 
Then, for PISO model $\lambda=60.692$ $M_{\odot}/\rm pc^3$, a value that is greater than the average value of the tension shown in Table \ref{TablePiso} but inside the interval marked by the standard deviation. 
For NFW model $\lambda=226.054$ $M_{\odot}/\rm pc^3$, this value is lower than the average value reported in Table \ref{TableNFW} and inside the range marked by the standard deviation. 
For Burkert model $\lambda=1.58\times 10^5$ $M_{\odot}/\rm pc^3$ this value is well above than the average value shown in Table \ref{TableBurkert}; a value outside the range marked by the standard deviation.

%%%%%%%%%%%%%%%%%%%%%%%%%%%%%%%%%%%%%%%%%%%%%%%
\section{Discussion and conclusions} \label{Disc}
%%%%%%%%%%%%%%%%%%%%%%%%%%%%%%%%%%%%%%%%%%%%%%%

We have presented in this paper, the effects coming from the presence of branes in galaxy rotation curves for 
three density profiles used to study the behavior of DM at galactic scales. 
With this in mind, we were given to the task of study a sample of 
high resolution measurements of rotation curves of galaxies without photometry\cite{deBlok/etal:2001} 
and a synthetic rotation curve built from 40 rotation curves of galaxies of magnitude around $M_I=-18.5$
fitting
 the values of $\rho_{s}$, $r_{s}$ and $\lambda$ through minimizing the $\chi^{2}_{red}$ value and we have compared with the standard results of $\rho_{s}$, $r_{s}$ for each DM density profile without branes. 
 The results for every observable in the three different profiles were summarized and compared in Tables \ref{TablePiso}-\ref{TableSynthetic}.

From here, it is possible to observe how the results show a weaker limit for the value of brane tension 
($\sim10^{-3}\; \rm eV^4-46$ eV$^4$) for the three models, in comparison with other astrophysical and cosmological studies\cite{Kapner:2006si,Alexeyev:2015gja,mk,gm,Sakstein:2014nfa,Kudoh:2003xz,Cavaglia:2002si,Holanda:2013doa,Barrow:2001pi,Brax:2003fv}; for example, Linares \emph{et al.}\cite{Linares:2015fsa} show that weaker values than $\lambda \simeq 10^{4}$ MeV$^{4}$, present anomalous behavior in the compactness of a dwarf star composed by a polytropic EoS, concluding that a wide region of their bound 
will show a non compactness stellar configuration, if it is applied to the study shown in\cite{Linares:2015fsa}.
 
It is important to notice that chosen a value of brane tension that not fulfill our bounds imposed through the paper, generate an anomalous behavior in the center of the galaxy which is characteristic of the model. Remarkable, for higher values of this bound, the modified rotation curves are in good agreement with 
the observed rotation curves of the sample that we use,
presenting only the distinctive features of each density profile: For example,
NFW dark matter density profile prefers lower values of the brane tension (on the average $\lambda \sim 0.73\times 10^{-3}$ eV$^4$), implying clear effects of the brane;
PISO dark matter
 case has an average value of $\lambda \sim 0.96\times 10^{-2}$ eV$^4$ and show relatively the maximum dispersion on the fitted values of the brane tension;
whereas Burkert DM density profile shows negligible brane effects, on the average $\lambda \sim 0.93$ eV$^4$ -- $46$ eV$^4$.

 In addition, it is important to discuss briefly the changes caused by the presence of branes in the problem of cusp/core. Notice that in this case the part that play a role is the effective density which is written in terms of brane corrections as: $\rho_{eff}=\rho(1-\rho/\lambda)$; it is notorious how the small perturbations alleviate the cusp problem which afflicts NFW, albeit the excessive presence of these terms could generates a negative effective density profile; also PISO and Burkert show modifications when $r\to0$ but does not pledge its core behavior while the brane tension only takes small values. In this way, the possibility of having a core behavior, help us to constraint the value of brane tension and still keep in the game the NFW profile.

Summarizing, it is really challenging to establish bounds in a dynamical systems like rotation curves in galaxies due to the
low densities found in the galactic medium, giving only a weaker limits in comparison with other studies in a most energetic systems. 
Our most important conclusion, is that despite the efforts, we think that it is not straightforward to do that the results fit with other astrophysical and cosmological studies, being impractical and not feasible to
find evidence of extra dimensions in galactic dynamics through the determination of the brane tension value, 
even more, we think that exist too much dispersion in the fitted values of the brane tension using this method for some DM density profile models. 
Also it is important to note that the value of the brane tension is strongly dependent of the characteristic of the galaxy studied, suggesting an average for the preferred value of the brane tension in each case. 
In addition, we notice that the effects of extra dimensions are stronger in the galactic core, suggesting that the NFW model is not appropriate in the search of constraints in brane theory due to the divergence in the center of the galaxy (see Eq.\ \eqref{comp}); PISO and Burkert could be good candidates to explore the galactic core in this framework; however it is necessary a more extensive study before we obtain a definitive conclusion.  

As a final note, we know that it is necessary to recollect more observational data to constraint 
the models or even give the final conclusion about extra dimensions (for or against), supporting the brane constraints shown through this paper with a more profound study of galactic dynamic or other tests like cosmological evidences presented in CMB anisotropies. 
However, this work is in progress and will be reported elsewhere.

%%%%%%%%%%%%%%%%%%%%%%%%%%%%%%%%%%%%%%%%%%%%%%%
\begin{acknowledgements}
MAG-A acknowledge support from SNI-M\'exico and CONACyT research fellow. Instituto Avanzado de Cosmolog\'ia (IAC) collaborations.
\end{acknowledgements}

\bibliography{librero1}

\newpage \
\newpage

%%%%%%%%%%%%%%%%%%%%%%%%%%%%%%%%%%%%%%%%%%%%%%
\begin{widetext}
\begin{table}
\begin{tabular}{llll||lllll}
\hline \hline 
& \multicolumn{3}{c||}{PISO} & \multicolumn{4}{c}{PISO + branes}  \\
\hline
Galaxy  &  $\rho_s$ ($M_{\odot}/\rm pc^3$) \; & $r_s$ (kpc) \; & $\chi_{red}^2$ 
& $\rho_s$ ($M_{\odot}/\rm pc^3$) \; & $r_s$ (kpc) \; & $\lambda$ ($M_{\odot}/ \rm pc^3$) 
\; & $\chi_{red}^2$
\\ \hline \hline
ESO 3050090 \quad	& 0.02732 	& 2.0911	 	& 0.052  \quad 	&  0.02738 	& 2.0889	 	& 13.148 		& 0.056 \\
ESO 0140040   	& 0.24930 	& 2.5588	 	& 0.216 		& 0.25012 	& 2.5546	 	& 77.901	 	& 0.271 \\
ESO 2060140		& 0.23308 	& 1.1638	 	& 0.115 		& 0.39715 	& 0.8863 		& 0.787	 	& 0.121 \\
ESO 3020120		& 0.05420 	& 1.8953	 	& 0.038 		& 0.11611 	& 1.2721	 	& 0.167 		& 0.031 \\
ESO 4250180		& 0.03019 	& 4.3977	 	& 0.114  		&  0.03028 	& 4.3915	 	& 10.597	 	& 0.143 \\
ESO 4880049  		& 0.10235 	& 1.6224	 	& 0.019 		& 0.10245 	& 1.6217	 	& 108.096 	& 0.022 \\
F 570\_V1 		& 0.21403 	&1.4723 		& 0.100 		& 0.21446 	& 1.4708	 	& 107.042 	& 0.124 \\
U 11454 			& 0.15115 	& 1.9263 		& 0.426  		& 0.15155 	& 1.9238	 	& 58.574	 	& 0.478 \\
U 11616 			& 0.19931 	& 1.4978	 	& 0.188  		& 0.45719 	& 0.9685	 	& 0.600	 	& 0.141 \\
U 11648 			& 0.10588 	& 1.9476	 	& 3.936 		& 0.11608 	& 1.8602	 	& 1.244 		& 4.167 \\
U 11748 			& 8.16533 	& 0.3674	 	& 6.101 		& 15.6937 	& 0.2652	 	& 19.680 		& 6.303 \\
U 11819 			& 0.08888 	& 2.9326	 	& 0.342  		& 0.15733 	& 2.1650	 	& 0.302	 	& 0.355
\\ \hline \hline
\end{tabular}
\caption{Parameters of the analyzed sample with PISO rotation velocity. From left to right, the columns read: name of the galaxy; central density in units of $M_{\odot}$/pc$^3$; central radius in kpc and the $\chi_{red}^2$ value.
And parameters of the analyzed sample with PISO+Branes rotation velocity:
central density in units of $M_{\odot}$/pc$^3$; central radius in kpc; brane tension in units of $M_{\odot}$/pc$^3$ and the $\chi_{red}^2$ value.
It is useful to have $\lambda$ in eV, the conversion from solar masses to eV is: $M_{\odot}/\rm pc^3\sim2.915\times10^{-4}eV^4$. 
Note that the average value of the fitted values of brane tension is
$\langle\lambda\rangle_{\rm PISO}\simeq33.178\;M_{\odot}/\rm pc^3$ with a standard deviation $\sigma_{\rm PISO}\simeq40.935\;M_{\odot}/\rm pc^3$.
}\label{TablePiso}
\end{table}
\end{widetext}
%%%%%%%%%%%%%%%%%%%%%%%%%%%%%%%%%%%%%%%%%%%%%%

\newpage \
\newpage \
\newpage

%%%%%%%%%%%%%%%%%%%%%%%%%%%%%%%%%%%%%%%%%%%%%%
\begin{widetext}
\begin{table}
\begin{tabular}{llll||llll}
\hline \hline
& \multicolumn{3}{c||}{NFW} &  \multicolumn{4}{c}{NFW + branes} \\
\hline
Galaxy & $\rho_s$ ($M_{\odot}/\rm pc^3$) \; & $r_s$ (kpc) \; & $\chi_{red}^2$ 
&  $\rho_s$ ($M_{\odot}/\rm pc^3$) \; & $r_s$ (kpc) \; & $\lambda$ ($M_{\odot}/ \rm pc^3$) 
\; & $\chi_{red}^2$
\\ \hline \hline
ESO 3050090	\quad	& 0.00003 	& 907.272 & 0.238 \quad 	&  0.00002 	& 1385.81	& 0.489 		& 0.164 \\
ESO 0140040   		& 0.02548 	& 16.148 	& 0.170 		& 0.03272 	& 14.182 	& 1.995 		& 0.104 \\
ESO 2060140			& 0.01989 	& 8.110 	& 0.455  		& 0.01997 	& 8.214 	& 1.640 		& 0.339 \\
ESO 3020120			& 0.00262 	& 19.720 	& 0.369 		& 0.00455 	& 13.763 	& 0.513 		& 0.191 \\
ESO 4250180			& 0.00052 	& 119.413 & 0.015 		& 0.00069 	& 94.037 	& 1.703 		& 0.011 \\
ESO 4880049 			& 0.00134 	& 54.811 	& 0.185 		& 0.00165 	& 45.696 	& 7.662 		& 0.141 \\
F 570\_V1 			& 0.01064 	& 14.707 	& 1.232 		& 0.01325 	& 12.799 	& 4.122 		& 1.140 \\
U 11454 				& 0.00797 	& 19.013 	& 3.208 		& 0.01731 	& 11.607 	& 1.175 		& 0.227 \\
U 11616 				& 0.01167 	& 13.811 	& 1.606 		& 0.02679 	& 8.221 	& 1.245 		& 0.384 \\
U 11648 				& 0.00404 	& 24.409 	& 0.990  		& 0.00404 	& 24.408 	& 4323.28 	& 1.045 \\
U 11748 				& 0.49834 	& 3.191 	& 3.782 		& 0.82773 	& 2.574 	& 9.232 		& 2.163 \\
U 11819 				& 0.00240 	& 52.691 	& 1.514 		& 0.01033 	& 18.102 	& 0.487 		& 0.690
\\ \hline \hline
\end{tabular}
\caption{Parameters of the analyzed sample with NFW rotation velocity. From left to right, the columns read: name of the galaxy; central density in units of $M_{\odot}$/pc$^3$; central radius in kpc and the $\chi_{red}^2$ value.
And properties and parameters of the analyzed sample with NFW+Branes rotation velocity: 
central density in units of $M_{\odot}$/pc$^3$; central radius in kpc; brane tension in units of $M_{\odot}$/pc$^3$ and the $\chi_{red}^2$ value.
Note that the
average value of the fitted values of brane tension is
$\langle\lambda\rangle_{\rm NFW}\simeq 362.795 \; M_{\odot}/\rm pc^3$ with a standard deviation $\sigma_{\rm NFW}\simeq 1194.134 \; M_{\odot}/\rm pc^3$. See the text for more details about the average without the outliers galaxies.
}\label{TableNFW}
\end{table}
\end{widetext}
%%%%%%%%%%%%%%%%%%%%%%%%%%%%%%%%%%%%%%%%%%%%%%

\newpage \
\newpage \
\newpage

%%%%%%%%%%%%%%%%%%%%%%%%%%%%%%%%%%%%%%%%%%%%%%
\begin{table}
\begin{tabular}{llll||llll}
\hline \hline
& \multicolumn{3}{c||}{Burkert} & \multicolumn{4}{c}{Burkert + branes} \\
\hline
Galaxy & $\rho_s$ ($M_{\odot}/\rm pc^3$) \; & $r_s$ (kpc) \; & $\chi_{red}^2$ 
& $\rho_s$ ($M_{\odot}/\rm pc^3$) \; & $r_s$ (kpc) \; & $\lambda$ ($M_{\odot}/ \rm pc^3$) 
\; & $\chi_{red}^2$
\\ 
\hline \hline
ESO 3050090	\quad	& 0.03069 & 3.376 & 0.045 		& 0.03069 & 3.376 & 3139.12 	& 0.049 \\
ESO 0140040			& 0.17489 & 5.973 & 0.667 		& 0.17490 & 5.973 & 4545.85 	& 0.834 \\
ESO 2060140			& 0.19000 & 2.480 & 0.209  		& 0.19002 & 2.480 & 5995.5 	& 0.228 \\
ESO 3020120			& 0.05668 & 3.372 & 0.007 \quad 	& 0.08724 & 2.798 & 0.174 	& 0.002 \\
ESO 4250180			& 0.03071 & 7.695 & 0.134 		& 0.03071 & 7.695 & 441.653 	& 0.167  \\
ESO 4880049 			& 0.11095 & 2.729 & 0.051 		& 0.11095 & 2.728 & 1682.83 	& 0.059 \\
F 570\_V1 			& 0.19128 & 2.391 & 0.683 		& 0.19129 & 2.931 & 5268.5 	& 0.854  \\
U 11454 				& 0.13899 & 3.705 & 1.090  		& 0.13900 & 3.705 & 1203.16 	& 1.227  \\
U 11616 				& 0.19596 & 2.751 & 0.359 		& 0.19598 & 2.751 & 985.585 	& 0.399  \\
U 11648 				& 0.08180 & 4.217 & 5.737 		& 0.08180 & 4.217 & 5465.98 	& 6.055 \\
U 11748 				& 1.43835 & 1.932 & 2.309 		& 2.56134 & 1.555 & 2.777 	& 2.429 \\
U 11819 				& 0.09828 & 4.843 & 0.204 		& 0.13497 & 4.189 & 0.364 	& 0.207
\\ \hline \hline
\end{tabular}
\caption{Parameters of the analyzed sample with Burkert rotation velocity. From left to right, the columns read: name of the galaxy; central density in units of $M_{\odot}$/pc$^3$; central radius in kpc and the $\chi_{red}^2$ value.
And parameters of the analyzed sample with Burkert+Branes rotation velocity: 
central density in units of $M_{\odot}$/pc$^3$; central radius in kpc; brane tension in units of $M_{\odot}$/pc$^3$ and the $\chi_{red}^2$ value.
Note that the
average value of the fitted values of brane tension is
$\langle\lambda\rangle_{\rm Burk}\simeq 2394.060 \;M_{\odot}/\rm pc^3$, $\sigma_{\rm Burk}\simeq2250.061\;M_{\odot}/\rm pc^3$. See the text for more details about the average without the outliers galaxies.
}\label{TableBurkert}
\end{table}
%%%%%%%%%%%%%%%%%%%%%%%%%%%%%%%%%%%%%%%%%%%%%%

\newpage \
\newpage \
\newpage

%%%%%%%%%%%%%%%%%%%%%%%%%%%%%%%%%%%%%%%%%%%%%%
\begin{table}
\begin{tabular}{llll||lllll}
\hline \hline 
%& \multicolumn{3}{c||}{PISO} & \multicolumn{4}{c}{PISO + branes}  \\
%\hline
DM  profile  &  $\rho_s$ ($M_{\odot}/\rm pc^3$) \; & $r_s$ (kpc) \; & $\chi_{red}^2$ 
& $\rho_s$ ($M_{\odot}/\rm pc^3$) \; & $r_s$ (kpc) \; & $\lambda$ ($M_{\odot}/\rm pc^3$) 
\; & $\chi_{red}^2$
\\ \hline \hline
PISO \quad	& 2.637 	& 0.968	 	& 0.447  \quad 	&  2.761 	& 0.945	 & 60.692 		& 0.511 \\
NFW   	& 0.0044 	& 244.27	 	& 6.019 		& 0.123	& 12.064	 	& 226.054	 	& 1.711 \\
Burkert		& 2.958 	& 1.571	 	& 0.321 		& 2.958 	& 1.571 		& $1.58\times 10^{5}$	& 0.367 
\\ \hline \hline
\end{tabular}
\caption{Parameters of the analyzed synthetic rotation curve with PISO, NFW and Burkert rotation velocity. From left to right, the columns read: name of the DM density profile; central density;
central radius and the $\chi_{red}^2$ value. And parameters of the analyzed synthetic rotation curve with DM profile+Branes rotation velocity: central density; central radius; brane tension and the $\chi_{red}^2$ value. 
}\label{TableSynthetic}
\end{table}
%%%%%%%%%%%%%%%%%%%%%%%%%%%%%%%%%%%%%%%%%%%%%%
\end{document}